\documentclass[peerreview]{IEEEtran}
\usepackage{amsthm}
\usepackage{amsmath}
\usepackage{mathrsfs}
\usepackage{amssymb} 
\usepackage{bbm,dsfont} 
\usepackage{multirow} 
\usepackage{units}
\usepackage{stmaryrd}   
\usepackage{verbatim}
\usepackage{enumerate} 
\usepackage[table]{xcolor}
\usepackage{slashbox} 
\usepackage{ wasysym }
\usepackage{tikz}
\usetikzlibrary{calc}
\usepackage{tabularx}

\usepackage{lipsum}
\usepackage{multicol}
\usepackage{float}

\usepackage{ifthen}

\usepackage{xcolor}
\usepackage{hyperref} 
\usepackage
{hyperref} 
\hypersetup{
    colorlinks,
    linkcolor={blue!80!black},
    citecolor={green!50!black},
    urlcolor={blue!80!black}
}

\usepackage{graphicx}
\graphicspath{{images/}}

\title{Entanglement-Assisted Capacity of Quantum Channels with Side Information}
\author{\IEEEauthorblockN{Uzi Pereg}\\
\IEEEauthorblockA{Department of Electrical Engineering, 
Technion, Haifa 32000, Israel.\\
Email: {\tt uzipereg@campus.technion.ac.il}
 }}

\usepackage[square,numbers]{natbib}
\definecolor{light-gray}{gray}{0.8}

\definecolor{dark-gray}{gray}{0.3}
\usepackage{accents}
\newlength{\dhatheight}

\newcommand{\bieee}{\begin{IEEEeqnarray}{rCl}}
\newcommand{\eieee}{\end{IEEEeqnarray}}
\newcommand{\prob}[1]{\Pr\left(#1\right)}

\renewcommand{\mathbbm}[1]{\text{\usefont{U}{bbm}{m}{n}#1}} 
\newcommand{\eps}{\varepsilon}

\newcommand{\norm}[1]{\left\lVert#1\right\rVert}
\newcommand{\trace}{\mathrm{Tr}}

\newcommand{\identity}{\mathbbm{1}}
\newcommand{\kb}[1]{ | #1 \rangle\langle #1 | } 

\newcommand{\ie}{\emph{i.e.} }
\newcommand{\eg}{\emph{e.g.} }
\newcommand{\etal}{\emph{et al.} }

\newcommand{\tR}{\widetilde{R}}

\newcommand{\hP}{\hat{P}}
\newcommand{\hM}{\hat{M}}

\newcommand{\Aset}{\mathcal{A}}

\newcommand{\Dset}{\mathcal{D}}
\newcommand{\Fset}{\mathcal{F}}

\newcommand{\Hset}{\mathcal{H}}

\newcommand{\Mset}{\mathcal{M}}

\newcommand{\Sset}{\mathcal{S}}

\newcommand{\Xset}{\mathcal{X}}
\newcommand{\Yset}{\mathcal{Y}}
\newcommand{\Zset}{\mathcal{Z}}
\newcommand{\Eset}{\mathcal{E}}

\theoremstyle{remark}	\newtheorem{theorem}{Theorem}
\theoremstyle{remark}	\newtheorem{lemma}[theorem]{Lemma}
\theoremstyle{remark}	\newtheorem{corollary}[theorem]{Corollary}
\theoremstyle{remark}	
\theoremstyle{remark} \newtheorem{definition}{Definition}
\theoremstyle{remark} \newtheorem{remark}{Remark}
\theoremstyle{remark} \newtheorem{example}{Example}

\newcommand{\pSpace}{\mathcal{P}}														

\newcommand{\tset}{\Aset^{\delta}}													
\newcommand{\Tset}{\mathcal{T}}												

\newcommand{\channel}{\mathcal{N}}

\newcommand{\inC}{\mathsf{C}}
\newcommand{\opC}{\mathbb{C}}
\newcommand{\opQ}{\mathbb{Q}}
\newcommand{\CclEA}{\opC_{E,\text{n-c}}(\channel)} 
\newcommand{\CqEA}{\opQ_{E,\text{n-c}}(\channel)} 

\newcommand{\CclEAnoSI}{\opC(\channel^{(0)})} 
\newcommand{\CqEAnoSI}{\opQ(\channel^{(0)})} 

\begin{document}
\maketitle

{}

\begin{abstract} 
Entanglement-assisted communication over a random-parameter quantum channel with either causal or non-causal channel 
side information (CSI) at the encoder is considered.
This describes a scenario where the quantum channel depends on the quantum state of the input environment. While Bob, the decoder,  has no access to this state, Alice, the transmitter, performs a sequence of projective measurements on her environment to encode her message. 
Dupuis \cite{Dupuis:08a,Dupuis:09c} established the entanglement-assisted capacity with non-causal CSI. Here, we establish characterization in the causal setting, and  also
give an alternative proof technique and further observations for the non-causal setting. 
\end{abstract}

\begin{IEEEkeywords}
Quantum information, Shannon theory, communication, 
channel capacity, state information, entanglement assistance.
\end{IEEEkeywords}

\section{Introduction}
A fundamental task in classical information theory is to determine the ultimate transmission rate of communication.
Shannon's channel coding theorem \cite{Shannon:48p} states that for a given noisy channel, with a transition probability function 
$p_{Y|X}$,  a vanishing probability of error is achievable as long  as the transmission rate is lower than the channel capacity, given by $C(p_{Y|X})=\max_{p_X} I(X;Y)$, where $I(X;Y)$ 
is the mutual information  between the channel input $X$ and 
output $Y$. For rates above the channel capacity, reliable communication cannot be accomplished.

Various classical settings of practical significance can be described by a channel $p_{Y|X,S}$ that depends on a random parameter $S$ when there is causal or non-causal channel side information (CSI) available at the encoder (see \eg \cite{Jafar:06p,KeshetSteinbergMerhav:07n,ChoudhuriKimMitra:13p} and references therein).
For example,    a cognitive radio in a wireless system may be aware of the channel state and network configuration \cite{GJMS:09p,Haykin:05p,BaruchShamaiVerdu:08c}, 
memory storage where the writer knows the fault locations \cite{HeegardElGamal:83p,KuznetsovTsybakov:74p}, 
and digital watermarking where the host data is treated as side information 
(see \eg \cite{ChenWornell:01p,SteinbergMerhav:01p,MoulinOsullivan:03p}).
The capacity with causal CSI is given by \cite{Shannon:58p}
 \begin{align}
C_{E,\text{caus}}(p_{Y|X,S})=\max_{p_{T}} I(T;Y)
\label{eq:CClcaual}
\end{align}
with $X=T(S)$, where $T:\Sset\rightarrow \Xset$ is called a \emph{Shannon strategy}
(see also \cite{KeshetSteinbergMerhav:07n,ChoudhuriKimMitra:13p}).
A channel with non-causal CSI is often referred to as the Gel'fand-Pinsker model \cite{GelfandPinsker:80p}. The capacity of this channel is given by
 \begin{align}
C_{E,\text{n-c}}(p_{Y|X,S})=\max_{p_{U,X|S}}[ I(U;Y)-I(U;S) ]
\label{eq:Cgp}
\end{align}
 where $U$ is an auxiliary random variable.

The 
field of quantum information is rapidly evolving in both practice and theory 
\cite{DowlingMilburn:03p,JKLGD:13p,BennettBrassard:14p,KEHZ:15p,BecerraFanMigdall:15p,YCLZRC:17p,LiaoGuoHuang:17p,ZDSZSG:17p}.
As technology 
approaches the atomic scale, 
we seem to be on the verge of the  ``Quantum Age" \cite{BouwmeesterZeilinger:00b,ImreGyongyosi:12b}.
Dynamics can sometimes be modeled by a noisy quantum channel, describing
 physical evolutions, density transformation,
discarding of sub-systems, quantum measurements, etc \cite{Kitaev:97b} \cite[Section 4.6]{Wilde:17b}.
%
Quantum information theory is the natural extension of classical information theory. Nevertheless, 
this generalization reveals astonishing phenomena with no parallel in classical communication \cite{GyongyosiImreNguyen:18p}. For example, 
 two quantum channels, each with zero quantum capacity, can have a nonzero quantum capacity
when used together \cite{SmithYard:08p}. This property is known as super-activation.

Communication through quantum channels can be separated into different categories.
In particular, one may consider a setting where Alice and Bob are provided with entanglement resources \cite{NielsenChuang:02b}.
 The entanglement-assisted capacity for transmission of  classical information over a quantum channel was fully characterized by Bennet \etal 
\cite{BennettShorSmolin:99p,BennettShorSmolin:02p}. Further work on entanglement-assisted communication
can be found \eg in \cite{Holevo:02p,HsiehDevetakWinter:08p,DevetakHarrowWinter:08p,Shirokov:12p,DattaHsieh:13p,WildeHsiehBabar:14p,
QianZhan:18p,AnshuJainWarsi:17a,CCVH:19a,AnshuJainWarsi:19p}.
 As for classical communication without entanglement between the encoder and the decoder,
the Holevo-Schumacher-Westmoreland (HSW) Theorem provides an asymptotic (``multi-letter")  formula for the capacity \cite{Holevo:98p,SchumacherWestmoreland:97p}, though calculation of such a formula is intractable in general. This is because the Holevo information is not necessarily additive \cite{Hastings:09p}. 
Shor has shown that the Holevo information is additive for the class of entanglement-breaking channels  \cite{Shor:02p}, in which case the HSW theorem provides a single-letter computable formula for the classical capacity. This class includes both classical-quantum channels and 
quantum-classical channels \cite[Section 4.6.7]{Wilde:17b}.
A similar difficulty occurs with transmission of quantum information over a quantum channel.
A multi-letter formula for the quantum capacity is given in \cite{BarnumNielsenSchumacher:98p,Loyd:97p,Shor:02l,Devetak:05p}, in terms of the coherent information. 
A computable 
formula is obtained in the special case where the channel is degradable 
\cite{DevetakShor:05p}.

The entanglement-assisted capacity of a quantum channel with non-causal CSI was determined by Dupuis
\cite{Dupuis:08a,Dupuis:09c}. 
Furthermore, Boche, Cai, and N\"{o}tzel \cite{BocheCaiNotzel:16p} addressed the classical-quantum channel  with CSI at the encoder without entanglement. 
The classical  capacity 
was determined given causal CSI, and a multi-letter formula was provided given 
non-causal CSI. Warsi and Coon \cite{WarsiCoon:17p} used an information-spectrum approach to derive  multi-letter bounds for a similar setting, where the side information has a limited rate.
 Luo and Devetak \cite{LuoDevetak:09p} considered channel simulation with source side information (SSI) at the 
decoder, and also solved the quantum generalization of the Wyner-Ziv problem \cite{WynerZiv:76p}.  Quantum data  compression with SSI  is also studied in \cite{DevetakWinter:03p,YardDevetak:09p,HsiehWatanabe:16p,DattaHircheWinter:19c,DattaHircheWinter:18a,
CHDH:19c, CHDH:18a} without entanglement-assitance. Compression with SSI given entanglement assistance was recently considered by 
Khanian and Winter  \cite{KhanianWinter:19c2,KhanianWinter:18a,KhanianWinter:19c,KhanianWinter:19a}.

In this paper, we consider a quantum channel  with either causal or non-causal CSI.
The motivation is as follows. Suppose that Alice wishes to send classical information to Bob through a (fully) quantum channel
$\channel_{SA\rightarrow B}$, where $A$ is the transmitter system, $B$ is the receiver system, and $S$ is the transmitter's environment, which affects the channel as well. Furthermore, suppose that Alice performs a sequence of projective measurements of the environment system $S$, hence the system is projected onto a particular vector $|s\rangle$ with probability $q(s)$. Using the measurement results, Alice encodes her message and sends her transmission through the channel. Whereas, Bob, who does not have access to the measurement results, ``sees" the average channel 
$\sum_s q(s) \channel^{(s)}_{A\rightarrow B}$, where $\channel^{(s)}_{A\rightarrow B}$ is the projection of the channel onto $|s\rangle$. 
Assuming Alice's  measurement projects onto orthogonal vectors, the environment system can be thought of as a classical random parameter 
$S\sim q(s)$. Therefore, we treat the quantum counterpart of the models in \cite{Shannon:58p} and \cite{GelfandPinsker:80p}, \ie 
a random-parameter quantum channel $\channel_{SA\rightarrow B}$ with CSI at the encoder.

We give a full characterization of the entanglement-assisted classical capacity and quantum capacity with causal CSI, 
and  also
give an alternative proof technique and further observations for the non-causal setting.
%
While Dupuis' analysis with non-causal CSI in \cite{Dupuis:08a,Dupuis:09c} is based on the decoupling approach for the transmission of quantum information (qubits),  we take a more direct approach.
In our analysis, we incorporate the classical binning technique \cite{HeegardElGamal:83p} into the quantum packing lemma \cite{HsiehDevetakWinter:08p}. Essentially, in the achievability proof, Alice performs classical compression of the parameter sequence, and then transmits both the classical message and the compressed representation using a random phase variation of the superdense coding protocol (see \eg \cite{HsiehDevetakWinter:08p,Wilde:17b}). 
The results are analogous to those in the classical case, although, as usual, the quantum analysis is  more involved.
As observed in  \cite{GelfandPinsker:80p,HeegardElGamal:83p}, the classical optimization (\ref{eq:Cgp}) can be restricted to mappings from $(U,S)$ to  $X$ that are deterministic. In analogy, we observe that optimization over isometric maps suffices for our problem.
With causal CSI, quantum operations are applied in a reversed order, and the Shannon strategy in (\ref{eq:CClcaual}) is replaced with a quantum channel.

\section{Definitions and Related Work}
We begin with basic definitions. 
\subsection{Notation, States, and Information Measures}
 We use the following notation conventions. 
Calligraphic letters $\Xset,\Yset,\Zset,...$ are used for finite sets.
Lowercase letters $x,y,z,\ldots$  represent constants and values of classical random variables, and uppercase letters $X,Y,Z,\ldots$ represent classical random variables.  
 The distribution of a  random variable $X$ is specified by a probability mass function (pmf) 
	$p_X(x)$ over a finite set $\Xset$. 
 We use $x^j=(x_1,x_{2},\ldots,x_j)$ to denote  a sequence of letters from $\Xset$. 
 A random sequence $X^n$ and its distribution $p_{X^n}(x^n)$ are defined accordingly. 
For a pair of integers $i$ and $j$, $1\leq i\leq j$, we write a discrete interval as $[i:j]=\{i,i+1,\ldots,j \}$. 
%

The state of a quantum system $A$ is given by a density operator $\rho$ on the Hilbert space $\Hset_A$.
A density operator is an Hermitian, positive semidefinite operator, with unit trace, \ie 
 $\rho^\dagger=\rho$, $\rho\succeq 0$, and $\trace(\rho)=1$.
The state is said to be pure if $\rho=\kb{\psi}$, for some vector $|\psi\rangle\in\Hset_A$, where
$\langle \psi |$ 
is the Hermitian conjugate of $|\psi\rangle$. 
In general, a density operator has a spectral decomposition of the following form,
\begin{align}
\rho=\sum_{z\in\Zset} p_Z(z) \kb{ \psi_z } 
\end{align}
where $\Zset=\{1,2,\ldots,|\Hset_A|\}$, $p_Z(z)$ is a probability distribution over $\Zset$, and $\{ |\psi_z\rangle \}_{z\in\Zset}$ forms an orthonormal basis of the Hilbert space $\Hset_A$.
 The density operator can thus be thought of as an average of pure states.
A measurement of a quantum system is any set of operators $\{\Lambda_j \}$ that forms a positive operator-valued measure (POVM), \ie
the operators are positive semi-definite and 
$\sum_j \Lambda_j=\identity$, where $\identity$ is the identity operator (see 
\cite[Definition 4.2.1]{Wilde:17b}). According to the Born rule, if the system is in state $\rho$, then the probability of the measurement outcome $j$ is given by $p_A(j)=\trace(\Lambda_j \rho)$.

Define the quantum entropy of the density operator $\rho$ as
\begin{align}
H(\rho) \triangleq& -\trace[ \rho\log(\rho) ]
\end{align}
which is the same as the Shannon entropy 
associated with the eigenvalues of $\rho$.
We may also consider the state of a pair of systems $A$ and $B$ on the tensor product $\Hset_A\otimes \Hset_B$ of the corresponding Hilbert spaces.
Given a bipartite state $\sigma_{AB}$, 
define the quantum mutual information by
\begin{align}
I(A;B)_\sigma=H(\sigma_A)+H(\sigma_B)-H(\sigma_{AB}) \,. 
\end{align} 
Furthermore, conditional quantum entropy and mutual information are defined by
$H(A|B)_{\sigma}=H(\sigma_{AB})-H(\sigma_B)$ and
$I(A;B|C)_{\sigma}=H(A|C)_\sigma+H(B|C)_\sigma-H(A,B|C)_\sigma$, respectively.

A pure bipartite state 
is called \emph{entangled} if it cannot be expressed as the tensor product 
of two states 
in $\Hset_A$ and $\Hset_B$. 
The maximally entangled state 
between two systems 
of dimension $D$ 
is defined by
$
| \Phi_{AB} \rangle = \frac{1}{\sqrt{D}} \sum_{j=0}^{D-1} |j\rangle_A\otimes |j\rangle_B 
$, where $\{ |j\rangle_A \}_{j=0}^{D-1}$ and $\{ |j\rangle_B \}_{j=0}^{D-1}$  
are respective orthonormal bases. 
Note that $I(A;B)_{\kb{\Phi}}=2\cdot \log(D)$.

\subsection{Quantum Channel}
\label{subsec:Qchannel}
A quantum channel maps a quantum state at the sender system to a quantum state at the receiver system. 
Here, we consider a channel that is governed by a random parameter with a particular distribution.
Formally, a random-parameter quantum channel  is defined as a   linear, completely positive, trace preserving map 
$
\channel_{S A\rightarrow B}  
$, 
corresponding to a quantum physical evolution. The channel parameter $S$ can also be thought of as a classical system at state 
\begin{align}
\rho_S=\sum_{s\in\Sset} q(s) \kb{ s }
\end{align} 
where $\{|s\rangle\}_{s\in\Sset}$ is an orthonormal basis of the Hilbert space $\Hset_S$. 
A quantum channel has a Kraus representation
\begin{align}
\channel_{S A\rightarrow B}(\rho)=\sum_j N_j \rho_{SA} N_j^\dagger 
\end{align}
for all $\rho_{SA}$, where the operators $N_j$ satisfy $\sum_j N_j^\dagger N_j=\identity$ \cite[Section 4.4.1]{Wilde:17b}.
The projection on $|s\rangle$ is then given  by
\begin{align}
\channel^{(s)}_{ A\rightarrow B}(\rho)=\sum_j N_j^{(s)} \rho N_j^{(s)\,\dagger}  
\end{align}
where $N_j^{(s)}\equiv \langle s| N_j |s\rangle$.
A quantum channel is called isometric if it can be expressed as
$
\channel_{S A\rightarrow B}(\rho)= N \rho_{SA} N^\dagger 
$ 
where the operator $N$ is an isometry, \ie $ N^\dagger N=\identity$ \cite[Section 4.6.3]{Wilde:17b}.

We assume that both the random parameter state and the quantum channel have a product form. That is, the state of 
the joint system $S^n=(S_1,\ldots,S_n)$ is $\rho_{S^n}=\rho_S^{\otimes n}$, and if the systems $A^n=(A_1,\ldots,A_n)$ are sent through $n$ channel uses, then the parameter-input state $\rho_{S^n}\otimes\rho_{A^n}$ undergoes the tensor product mapping
$
\channel_{S^n A^n\rightarrow B^n}\equiv  \channel_{S A\rightarrow B}^{\otimes n} 
$. 
Therefore, without CSI, the input-output relation is
\begin{align}
\rho_{B^n}= \sum_{s^n\in\Sset^n} q^n(s^n) \channel^{(s^n)}_{A^n\rightarrow B^n } (\rho_{A^n})
= \left(\sum_{s\in\Sset} q(s) \channel^{(s)}_{A\rightarrow B} \right)^{\otimes n} (\rho_{A^n}) 
\label{eq:QinOutn}
\end{align}
where $q^n(s^n)=\prod_{i=1}^n q(s_i)$ is the joint distribution of the parameter sequence and
$\channel^{(s^n)}_{A^n\rightarrow B^n}=\channel^{(s_1)}_{A\rightarrow B}\otimes\cdots\otimes\channel^{(s_n)}_{A\rightarrow B} $. 
The sender and the receiver are often referred to as Alice and Bob. 


\begin{figure}
\begin{center}
\includegraphics[scale=0.5,trim={0 5cm 1cm 5cm},clip]{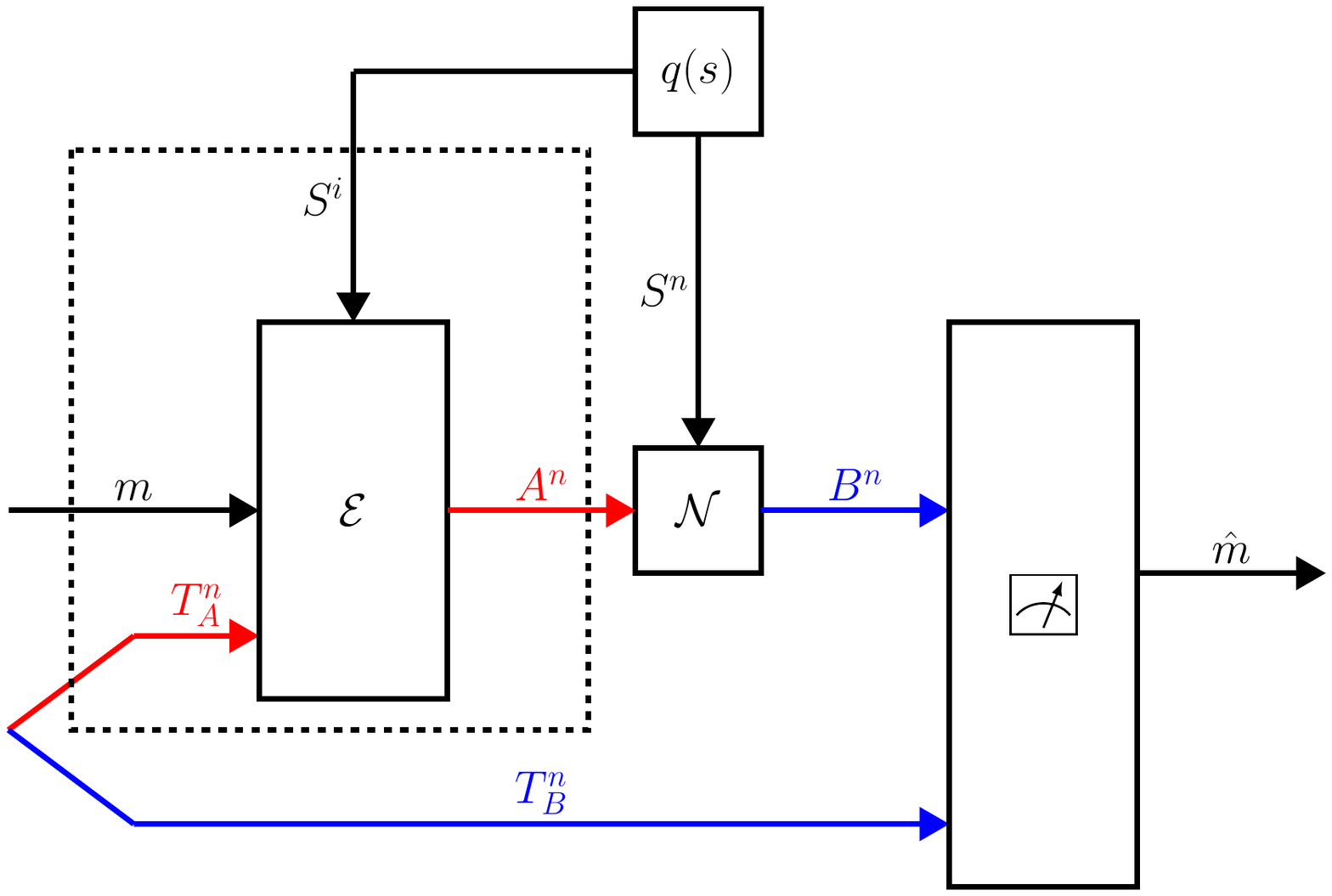} 
\end{center}
\caption{
Coding for a quantum channel $\channel_{SA\rightarrow B}$ that depends on a random parameter $S\sim q(s)$, with causal side information at the encoder. The quantum systems of Alice and Bob are marked in red and blue, respectively.
In particular, the systems inside the dashed-line rectangle are only available at the encoder.
 Alice chooses a classical message $m$. At time $i$, given the parameter sequence $s^i$, 
she applies the encoding channel $\Eset^{m,s^i}_{T_{A,i}\rightarrow A_i}$ to her share of the entangled state $\Psi_{T_{A,i},T_{B,i}}$, and then transmits the system $A_i$ over the quantum channel  $\channel_{SA\rightarrow B}$. 
 Bob receives the channel output systems $B^n$, combines them with the entangled system $T_B^n$, and performs a measurement. The outcome is the estimated message $\hat{m}$. 
}
\label{fig:EAsiCode}
\end{figure}

\subsection{Coding}
\label{subsec:Mcoding}
We define 
a  code to transmit classical information provided that the encoder and the decoder share unlimited entanglement. 
The entangled system pairs  are denoted by $(T_A^n,T_B^n)=(T_{A,i},T_{B,i})_{i=1}^n$.
With causal CSI, Alice knows the sequence of past and present random parameters, $S_1,\ldots,S_i$, at $i\in [1:n]$.

\begin{definition} 
\label{def:EAcapacity}
A $(2^{nR},n)$ entanglement-assisted classical code with causal CSI at the encoder consists of the following:   
a message set $[1:2^{nR}]$, where $2^{nR}$ is assumed to be an integer, a pure entangled state $\Psi_{T_{A}^n,T_{B}^n}$, 
  a sequence of $n$ encoding maps (channels) $\Eset^{m,s^i}_{T_{A,i}\rightarrow A_i}$, $m\in [1:2^{nR}]$, $s^i\in\Sset^i$, for $i\in [1:n]$, and a decoding POVM $\{ \Lambda^m_{B^n T_B^n}  \}_{m\in [1:2^{nR}]}$.
We denote the code by $(\Eset,\Psi,\Lambda)$.

The communication scheme is depicted in Figure~\ref{fig:EAsiCode}.  
The sender Alice has the systems $T_A^n,A^n$ and the receiver Bob has the systems $T_B^n,B^n$, where $T_A^n$ and $T_B^n$ are entangled. Alice chooses a classical message $m\in [1:2^{nR}]$. At time $i\in [1:n]$, given the sequence of past and present parameters
 $s^i\in\Sset^i$,  she applies the encoding channel $\Eset^{m,s^i}_{T_{A,i}\rightarrow A_i}$ to her share of the entangled state 
$\Psi_{T_{A,i},T_{B,i}}$, and then transmits the system $A_i$ over the channel. 
In other words, Alice uses an encoding channel $\overline{\Eset}^{m,s^n}_{T_A^n\rightarrow A^n}$ of the following form,
\begin{align}
\overline{\Eset}^{m,s^n}\triangleq \Eset^{m,s_1}\otimes\Eset^{m,s_1,s_2}\otimes\cdots\otimes \Eset^{m,s^n} \,,  
\end{align}
and transmits the systems $A^n$ over $n$ channel uses of $\channel_{SA\rightarrow B}$.

 Bob receives the channel output systems $B^n$, combines them with the entangled system $T_B^n$, and performs the POVM
 $\{ \Lambda^m_{B^n T_B^n}  \}_{m\in [1:2^{nR}]}$. The conditional probability of error, given that the message $m$ was sent, is given by 
\begin{align}
P_{e|m}^{(n)}(\Eset,\Psi,\Lambda)= \sum_{s^n\in\Sset^n} q^n(s^n)
\trace\Big[ (\identity-\Lambda^m_{B^n T_B}) 
(\channel^{(s^n)}_{A^n\rightarrow B^n}\otimes\identity)
(\overline{\Eset}^{m,s^n}\otimes\identity) (\Psi_{T_A^n,T_B^n})
 \Big] \,.
\end{align}
A $(2^{nR},n,\eps)$ entanglement-assisted classical 
code satisfies 
$
P_{e|m}^{(n)}(\Eset,\Psi,\Lambda)\leq\eps $ 
for all $m\in [1:2^{nR}]$. 
%
A rate $R>0$ is called achievable  if for every $\eps>0$ and sufficiently large $n$, there exists a $(2^{nR},n,\eps)$
code. The entanglement-assisted classical capacity $\opC_{E,\text{caus}}(\channel)$ is defined as the supremum of  achievable rates.
\end{definition}
Next, we give a definition of an entanglement-assisted quantum code.
A 
more general definition can be found in \cite{Wilde:17b}.
\begin{definition} 
\label{def:QEAcapacity}
A $(2^{nQ},n)$ entanglement-assisted quantum code with causal CSI consists of the following;  
A quantum state $\rho_{M}^{\otimes n}$, where $M$ is a system of dimension $2^{nQ}$;  
 a pure entangled state $\Psi_{T_A,T_B}$,
 a sequence of $n$ encoding channels $\Eset^{s^i}_{T_{A,i} M\rightarrow A_i}$, and a decoding channel 
$ \Dset_{B^n T_B^n\rightarrow \hM}  $.
 
The sender Alice has the systems $T_A^n,M,A^n$ and the receiver Bob has the systems $T_B^n,B^n,\hM$, where $T_A^n$ and $T_B^n$ are entangled.
Alice encodes the state $\rho_{M}$ by applying the encoding channel 
$\overline{\Eset}_{M,S^n,T_{A}^n \rightarrow A^n}$ to $\rho_{M}^{\otimes n}$ and to her share of the entangled state $\Psi_{T_{A}^n,T_{B}^n}$, where $\overline{\Eset}_{M,S^n,T_{A}^n \rightarrow A^n}=\bigotimes_{i=1}^n \Eset_{M,S^i,T_{A,i} \rightarrow A_i}$, and transmits the system $A^n$ over $n$ channel uses of 
$\channel_{SA\rightarrow B}$. Bob receives the channel output systems $B^n$, combines them with the entangled system $T_B$, and applies the decoding channel $ \Dset_{B^n T_B^n\rightarrow \hM}  $. 
The code 
is said to be a $(2^{nQ},n,\eps)$ entanglement-assisted quantum code if the trace distance between the original state and the resulting state at the receiver is bounded by
\begin{align}
\frac{1}{2}\norm{\rho_{M}-\Dset\left( \sum_{s^n\in\Sset^n} q^n(s^n) (\channel^{(s^n)}_{A^n\rightarrow B^n}\otimes\identity)
 (\overline{\Eset}^{s^n} \otimes\identity)      \left( \rho_{M}^{\otimes n}\otimes\Psi_{T_A^n,T_B^n} \right) \right) }_1
\leq \eps \,,
\end{align}
where $\norm{\cdot}_1$ denotes the trace norm.
A positive number $Q>0$ is said to be an achievable rate  if for every $\eps>0$ and sufficiently large $n$, there exists a $(2^{nQ},n,\eps)$
code. The entanglement-assisted quantum capacity $\opQ_{E,\text{caus}}$ is defined as the supremum of  achievable rates.
\end{definition}

We also discuss the non-causal setting, where Alice has the parameter sequence $S^n$ a priori, and can thus applies any encoding channel $\overline{\Eset}_{M,S^n,T_{A}^n \rightarrow A^n}$.
In addition, we consider the case where there is  CSI at the decoder, \ie when Bob receives both $B^n$ and $S^n$,  and performs a POVM $\{ \Lambda^m_{B^n S^n T_B}  \}_{m\in [1:2^{nR}]}$. We note that for the decoder, causality is insignificant.
We use the respective subscripts `$E$', `$D$' or `$ED$' to indicate that CSI is available at either the encoder, the decoder, or both,  and the subscripts `caus' or `n-c' to indicate whether CSI is available at the encoder in a causal or non-causal manner, respectively.
The notation is summarized in the table in Figure~\ref{table:LcapacityNotation}. 

\begin{figure*}[htb]	
\begin{center}
\hspace{-2cm}
\begin{tabular}{l|ccccc}
\backslashbox{Capacity}{CSI} 
 								  & $\;$ none $\;$ &	encoder $\;$ &  decoder 	 & encoder$+$decoder $\quad$  &	\textcolor{blue!80!black}{encoder (causal)} 
								 \\	[0.2cm]   \hline \\			[-0.2cm]
Classical					&	$\opC(\channel)$
									& $\CclEA$ 	
									& $\opC_D(\channel)$
									&	$\opC_{ED,\text{n-c}}(\channel)$
									& \textcolor{blue!80!black}{$\opC_{E,\text{caus}}(\channel)$}									
																																				\\[0.3cm] 
Quantum						& $\opQ(\channel)$
									&	$\CqEA$	
									&	$\opQ_D(\channel)$	
									& $\opQ_{ED,\text{n-c}}(\channel)$ 					
									& \textcolor{blue!80!black}{$\opQ_{E,\text{caus}}(\channel)$}										
\end{tabular}
\end{center}
  \caption{Notation of channel capacities with and without CSI.
	The columns correspond to the location where CSI is available, and the rows indicate the type of information capacity  -- classical or quantum. 
	}
\label{table:LcapacityNotation}
\end{figure*}

\subsection{Related Work}
\label{subsec:Previous}
We briefly review known results for a quantum channel that does not depend on a random parameter, \ie $\channel^{(s)}_{A\rightarrow B}=\channel^{(0)}_{A\rightarrow B}$ for $s\in\Sset$.
Define 
\begin{align}
\inC(\channel^{(0)})\triangleq \max_{|\phi\rangle_{AA'}} I(A;B)_\rho 
\end{align}
 with $\rho_{AB}\equiv (\identity\otimes\channel^{(0)})( \kb{ \phi }_{AA'})$. Next, we give the respective capacity theorems for the entanglement-assisted classical capacity and the entanglement-assisted quantum capacity.
\begin{theorem} [see {\cite{BennettShorSmolin:99p,BennettShorSmolin:02p}}]
\label{theo:CeaNoSI}
The entanglement-assisted classical capacity of a quantum channel $\channel_{A\rightarrow B}^{(0)}$ is given by 
\begin{align}
\CclEAnoSI=\inC(\channel^{(0)}) \,.
\end{align}
\end{theorem}
Given an unlimited supply of entanglement, the teleportation protocol can send a qubit using two classical bits, while
 the super-dense coding protocol can send two classical bits using one qubit \cite{NielsenChuang:02b}.
This implies the following.
\begin{corollary} [see {\cite{BennettShorSmolin:99p,BennettShorSmolin:02p}}]
The entanglement-assisted quantum capacity of a quantum channel $\channel_{A\rightarrow B}^{(0)}$ is given by 
\begin{align}
\CqEAnoSI=\frac{1}{2}\inC(\channel^{(0)}) \,.
\end{align}
\end{corollary}

\begin{remark}
\label{rem:noSI}
We  note that the setting of a random-parameter quantum channel $\channel_{S A\rightarrow B }$ \emph{without} side information  is equivalent to that of a channel that does not depend on a state, with 
$
\channel_{A\rightarrow B}^{(0)}=\sum_{s\in\Sset} q(s) \channel_{A\rightarrow B}^{(s)}
$ 
(see (\ref{eq:QinOutn})). On the other hand, with side information at the encoder, this equivalence does not hold, as the channel input is correlated with the parameter sequence.
\end{remark}

\section{Information Theoretic Tools}
To derive our results, we use the quantum version of the method of types properties and techniques. The basic definitions and lemmas that are used in this paper are given below.

\subsection{Classical Types}
The type of a classical sequence $x^n$ is defined as the empirical distribution $\hP_{x^n}(a)=N(a|x^n)/n$ for $a\in\Xset$, where $N(a|x^n)$ is the number of occurrences of the symbol $a$ in the sequence $x^n$. The set of all types over $\Xset$ is then denoted by 
$\pSpace_n(\Xset)$.
The type class associated with a type  $\hP\in \pSpace_n(\Xset)$ is defined as the set of sequences of that type, 
\ie
\begin{align} 
\Tset(\hP)\equiv\left\{ x^n\in\Xset^n \,:\; \hP_{x^n}=\hP  \right\} \,.
\end{align}
For a pair of sequences $x^n$ and $y^n$, we give similar definitions in terms of the joint type $\hP_{x^n,y^n}(a,b)=N(a,b|x^n,y^n)/n$ for $a\in\Xset$, $b\in\Yset$, where $N(a,b|x^n,y^n)$ is the number of occurrences of the symbol pair $(a,b)$ in the sequence 
$(x_i,y_i)_{i=1}^n$. Given a sequence $y^n\in \Yset^n$, we further define the conditional type $\hP_{x^n|y^n}(a|b)=N(a,b|x^n,y^n)/N(b|y^n)$
and the conditional type class  
\begin{align} 
\Tset(\hP|y^n)\equiv\left\{ x^n\in\Xset^n \,:\; \hP_{x^n,y^n}(a,b)=\hP_{y^n}(b)\hP(a|b) \right\} \,.
\end{align}

Given a probability distribution $p_X\in\pSpace(\Xset)$,  the $\delta$-typical set  is defined as
\begin{align}
\tset(p_X)\equiv \bigg\{ x^n\in\Xset^n \,:\; 
\left| \hP_{x^n}(a) - p_X(a) \right|\leq\delta \quad\text{if $\, p_X(a)>0$}&  \nonumber\\ 
 \hP_{x^n}(a)=0 \quad\text{if $\, p_X(a)=0$} &, \;\text{$\forall$ $a\in\Xset$} \bigg\}
\end{align}

The covering lemma is a powerful tool in classical information theory \cite{CsiszarKorner:82b}. 
\begin{lemma}[Classical Covering Lemma {\cite{CsiszarKorner:82b}\cite[Lemma 3.3]{ElGamalKim:11b}}]
\label{lemm:covering}
Let $X^n\sim \prod_{i=1}^n p_X(x_i)$, $\delta>0$, and let $Z^n(m)$, $m\in [1: 2^{nR}]$, be independent random sequences distributed according to $\prod_{i=1}^n p_Z(z_i)$. Suppose that the sequence $X^n$ is pairwise independent of the sequences $Z^n(m)$, $m\in [1:2^{nR}]$. Then,
\begin{align}
\prob{ (Z^n(m),X^n)\notin\tset(p_{Z,X}) \,\text{for all $m\in [1: 2^{nR}]$}  }\leq \exp\left( -2^{n(R- I(Z;X)-\eps_n(\delta )} \right)
\end{align}
where $\eps_n(\delta)$ tends to zero as $n\rightarrow\infty$ and $\delta\rightarrow 0$.
\end{lemma}
Let $X^n\sim \prod_{i=1}^n p_X(x_i)$ be an information source sequence,  encoded by an index $m$ at compression rate $R$.
Based on the covering lemma above, as long as the compression rate is higher than $I(Z;X)$,
a set of random codewords, $ Z^n(m)\sim \prod_{i=1}^n p_Z(z_i) $, contains with high probability at least one sequence that is jointly typical with the source sequence.

Though originally stated in the context of lossy source coding, the classical covering lemma is useful in a variety of scenarios
\cite{ElGamalKim:11b}, including the random-parameter channel with non-causal CSI. In this case, the parameter sequence 
$S^n\sim \prod_{i=1}^n q(s_i)$ plays the role of the ``source sequence".

\subsection{Quantum Typical Subspaces}
Moving to the quantum method of types, 
suppose that the state of a system is generated from an ensemble $\{ p_X(x), |x\rangle \}_{x\in\Xset}$, hence, the average density operator is
\begin{align}
\rho=\sum_{x\in\Xset} p_X(x) \kb{x} \,.
\end{align}
Consider the  subspace  spanned by the vectors $| x^n \rangle$, $x^n\in\Tset(\hP)$, for a given type
$\hP\in\pSpace_n(\Xset)$. 
Then, 
the projector onto the subspace 
is given by
\begin{align}
\Pi_{A^n}(\hP)\equiv \sum_{x^n\in\Tset(\hP)} \kb{ x^n } \,.
\label{eq:ProjType}
\end{align}
Note that the dimension of  the subspace of type class $\hP$ is given by 
$\trace(\Pi_{A^n}(\hP))=|\Tset(\hP)|$. By classical type properties \cite[Lemma 2.3]{CsiszarKorner:82b} (see also  \cite[Property 15.3.2]{Wilde:17b}),
\begin{align}
(n+1)^{|\Xset|} 2^{nH(\rho)} \leq    \trace(\Pi_{A^n}(\hP))  \leq  2^{nH(\rho)} \,.
\label{eq:Tsize}
\end{align}

The projector onto the $\delta$-typical subspace is defined as
\begin{align}
\Pi^\delta(\rho)\equiv \sum_{x^n\in\tset(p_X)} \kb{ x^n } \,.
\end{align}
Based on \cite{Schumacher:95p} \cite[Theorem 12.5]{NielsenChuang:02b}, for every $\eps,\delta>0$ and sufficiently large $n$, the $\delta$-typical projector satisfies
\begin{align}
\trace( \Pi^\delta(\rho) \rho^{\otimes n} )\geq& 1-\eps  \label{eq:UnitT} \\
 2^{-n(H(\rho)+c\delta)} \Pi^\delta(\rho) \preceq& \,\Pi^\delta(\rho) \,\rho^{\otimes n}\, \Pi^\delta(\rho) \,
\preceq 2^{-n(H(\rho)-c\delta)}
\label{eq:rhonProjIneq}
\\
\trace( \Pi^\delta(\rho))\leq& 2^{n(H(\rho)+c\delta)} \label{eq:Pidim}
\end{align}
where $c>0$ is a constant.

To prove achievability for Theorem~\ref{theo:CeaNoSI} above, one may invoke the quantum packing lemma \cite{HsiehDevetakWinter:08p,Wilde:17b}.
Suppose that Alice employs a quantum codebook that consists  of $2^{nR}$ ``codewords" $x(m)$, $m\in [1:2^{nR}]$, by which she chooses a state from an ensemble $\{\rho_x \}_{x\in\Xset}$. The proof is based on random codebook generation, where the codewords are drawn at random according to an input distribution $p_X(x)$. To recover the transmitted message, Bob may perform the square-root measurement \cite{Holevo:98p,SchumacherWestmoreland:97p} using
a code projector $\Pi$ and codeword projectors $\Pi_x$, $x\in\Xset$, which project onto subspaces of the Hilbert space $\Hset$. 

%
The lemma below is a  simplified, less general, version of the quantum packing lemma by Hsieh, Devetak, and Winter \cite{HsiehDevetakWinter:08p}.
\begin{lemma}[Quantum Packing Lemma {\cite[Lemma 2]{HsiehDevetakWinter:08p}}]
\label{lemm:Qpacking}
Let $\sigma_{AB}$ be a joint state on the product Hilbert space $\Hset_A\otimes\Hset_B$, such that
\begin{align}
\sigma_B=\sum_{x\in\Xset} p_X(x) \rho_x
\end{align}
where $\{ p_X(x), \rho_x \}_{x\in\Xset}$ is a given random ensemble on $\Hset_B$. Furthermore,
suppose that there is  a code projector $\Pi$ and codeword projectors $\Pi_x$, $x\in\Xset$, that satisfy the following
\begin{align}
\trace(\Pi\rho_x)\geq&\, 1-\alpha \\
\trace(\Pi_x\rho_x)\geq&\, 1-\alpha \\
\trace(\Pi_x)\leq&\, 2^{n(H(\sigma_{A,B})+\alpha)}\\
\Pi \sigma_A \Pi \preceq&\, 2^{-n(H(\sigma_A)+H(\sigma_B)-\alpha)} \Pi 
\end{align}
for some $\alpha>0$.
Then, there exist codewords $x(m)$, $m\in [1:2^{nR}]$, and  a POVM $\{ \Lambda_m \}_{m\in [1:2^{nR}]}$ such that 
\begin{align}
  \trace\left( \Lambda_m \rho_{x(m)} \right)  \geq 1-2^{-n[ I(A;B)_\sigma-R-\eps_n(\alpha)]}
\end{align}
for all 
$m\in [1:2^{nR}]$, where $\eps_n(\alpha)$ tends to zero as $n\rightarrow\infty$ and $\alpha\rightarrow 0$. 
\end{lemma}
In our analysis, where there is non-causal CSI at the encoder, we  apply the packing lemma such that the quantum ensemble encodes both the message $m$ and a compressed representation of the parameter sequence $s^n$.

\section{Main Results}
We give our results on the random-parameter quantum channel $\channel_{S A\rightarrow B}$ with  CSI at the encoder.
\subsection{Causal Side Information at the Encoder}
We begin with our main result on the random-parameter quantum channel with causal CSI.
Define 
\begin{align}
\label{eq:inCeaCausal}
\inC_{\text{caus}}(\channel)\triangleq \max_{ \theta_{KA'} \,,\; \Fset_{K\rightarrow A}^{(s)} } I(K;B)_\omega 
\end{align}
where the maximization is 
over the quantum state $ \theta_{KA'}$ and the set of quantum channels
$\{ \Fset^{(s)}_{K\rightarrow A} \}_{s\in\Sset}$, with
\begin{align}
\label{eq:StateMax}
&\omega^s_{AA'}=  (\Fset^{(s)} \otimes \identity   )(\theta_{KA'}) \\
&\omega_{ASA'}=\sum_{s\in\Sset} q(s) \kb{ s } \otimes \omega^s_{AA'}  \\
&\omega_{AB}= (\identity\otimes \channel)(\omega_{ASA'}) \,.
\end{align}
Before we state the capacity theorem, we give the following lemma.
\begin{lemma}
\label{lemm:pureCeaCausal}
The maximization in (\ref{eq:inCeaCausal}) can be restricted to pure states $ \theta_{KA'}= \kb{ \xi_{KA'} } $.
\end{lemma}
 Lemma~\ref{lemm:pureCeaCausal} follows by state purification \cite[Exercise 13.4.4]{Wilde:17b}. The proof is given in 
Appendix~\ref{app:pureCeaCausal}.
Now, we give our main result.
\begin{theorem}
\label{theo:mainC}
The entanglement-assisted classical capacity of the random-parameter quantum channel $\channel_{S A\rightarrow B}$ with causal CSI at the encoder is given by 
\begin{align}
\opC_{\text{caus}}(\channel)=\inC_{\text{caus}}(\channel) \,.
\end{align}
\end{theorem}
The proof of Theorem~\ref{theo:mainC} is given in Appendix~\ref{app:mainC}.
To prove achievability, we apply the 
random coding techniques from \cite{BennettShorSmolin:99p,BennettShorSmolin:02p} to the virtual channel $\mathcal{M}_{K\rightarrow B}$, defined by 
\begin{align}
\mathcal{M}(\rho_K)=\sum_{s\in\Sset} q(s) \channel^{(s)}\left( \Fset^{(s)}(\rho_K)  \right) \,.
\end{align}

As without side information, a qubit is exchangeable with two classical bits, given unlimited entanglement. This follows by applying 
the teleportation protocol and the super-dense coding protocol (see \cite[Sections 1.3.7, 2.3]{NielsenChuang:02b} and also
\cite[Chapter 6]{Wilde:17b}). As a consequence, we can characterize the entanglement-assisted quantum capacity as well.
\begin{theorem}
\label{theo:QmainC}
The entanglement-assisted quantum capacity of the random-parameter quantum channel $\channel_{S A\rightarrow B}$ with causal CSI at the encoder is given by 
\begin{align}
\opQ_{\text{caus}}(\channel)=\frac{1}{2}\inC_{\text{caus}}(\channel) \,.
\end{align}
\end{theorem}

\subsection{Non-Causal Side Information at the Encoder}
The entanglement-assisted capacity of a quantum channel with non-causal CSI was determined by Dupuis
\cite{Dupuis:08a,Dupuis:09c}. Here, we use an alternative proof approach, which yields an equivalent formulation and further observations.
Define 
\begin{align}
\label{eq:inCea}
\inC_{E,\text{n-c}}(\channel)\triangleq \max_{ \theta_{KA'} \,,\; \Fset_{K\rightarrow A}^{(s)} } \left[ I(A;B)_\omega-I(A;S)_\omega \right] 
\end{align}
where the maximization is over the quantum state $ \theta_{KA'}$ and the set of quantum channels
$\{ \Fset^{(s)}_{K\rightarrow A} \}_{s\in\Sset}$, with
\begin{align}
&\omega^s_{AA'}=  (\Fset^{(s)} \otimes \identity   )(\theta_{KA'}) \\
&\omega_{ASA'}=\sum_{s\in\Sset} q(s) \kb{ s } \otimes \omega^s_{AA'}  \\
&\omega_{AB}= (\identity\otimes \channel)(\omega_{ASA'}) \,.
\end{align}
Before we state the capacity theorem, we give the following lemma.
\begin{lemma}
\label{lemm:pureCea}
The maximization in (\ref{eq:inCea}) can be restricted to pure states $ \theta_{KA'}= \kb{ \xi_{KA'} } $ and isometric channels
$\Fset^{(s)}_{K\rightarrow A}(\rho_A)=F^{(s)}\rho_A F^{(s)\,\dagger}  $.
\end{lemma}
The proof of Lemma~\ref{lemm:pureCea} is given in Appendix~\ref{app:pureCea}, using state purification and isomeric channel extension.
%
%
Not only Lemma~\ref{lemm:pureCea} simplifies the calculation of the formula in (\ref{eq:inCea}), but it will also be useful in our proof for the theorem below. 
\begin{theorem}[also in \cite{Dupuis:08a,Dupuis:09c}]
\label{theo:mainNC}
The entanglement-assisted classical capacity of the random-parameter quantum channel $\channel_{S A\rightarrow B}$ with non-causal CSI at the encoder is given by 
\begin{align}
\CclEA=\inC_{E,\text{n-c}}(\channel) \,.
\end{align}
\end{theorem}
The proof of Theorem~\ref{theo:mainNC} is given in Appendix~\ref{app:mainNC}.
As we explained before, given unlimited entanglement, a qubit is exchangeable with two classical bits, implying the following.
\begin{theorem}[also in \cite{Dupuis:08a,Dupuis:09c}]
\label{theo:QmainNC}
The entanglement-assisted quantum capacity of the random-parameter quantum channel $\channel_{S A\rightarrow B}$ with non-causal CSI at the encoder is given by 
\begin{align}
\CqEA=\frac{1}{2}\inC_{E,\text{n-c}}(\channel) \,.
\end{align}
\end{theorem}
In \cite{Dupuis:08a,Dupuis:09c}, Dupuis applied the decoupling approach to prove Theorem~\ref{theo:QmainNC},  and then, obtained the classical capacity theorem, Theorem~\ref{theo:mainNC}, as a consequence. The decoupling approach shows that qubits can be transmitted by decoupling between the encoder's reference system and the output system. Here, we have taken a more direct approach and devised a coding scheme for the transmission of classical information.

\subsection{Side Information at the Decoder}
In this subsection, we consider a random-parameter quantum channel $\channel_{SA\rightarrow B}$ with CSI at  the decoder. That is, Bob receives both $B^n$ and $S^n$,  and performs a POVM $\{ \Lambda^m_{B^n S^n T_B}  \}_{m\in [1:2^{nR}]}$.
The results in this subsection are a straightforward consequence of the results above. 

First, suppose that only Bob is aware of the channel parameter sequence, and define
\begin{align}
\inC_D(\channel)=\max_{|\phi\rangle_{AA'}} I(A;B|S)_\rho 
\end{align}
  with 
\begin{align}	
	\rho_{SAB}\equiv \sum_{s\in\Sset} q(s) \kb{ s } \otimes (\identity\otimes\channel^{(s)})( \kb{ \phi }_{AA'}) \,.
	\label{eq:rhoSAB}
	\end{align}
\begin{corollary}
\label{coro:mainNCd}
The entanglement-assisted classical capacity of the random-parameter quantum channel $\channel_{S A\rightarrow B}$ with  CSI at the decoder is given by 
\begin{align}
\opC_{D}(\channel)=\inC_{D}(\channel) 
\end{align}
and the entanglement-assisted quantum capacity is given by 
$
\opQ_{D}(\channel)=\frac{1}{2}\inC_{D}(\channel) 
$. 
\end{corollary}
Corollary~\ref{coro:mainNCd} is a straightforward consequence of Theorem~\ref{theo:CeaNoSI}, 
following the observation that the channel parameter $S$ can be thought of as part of the output system in this setting.
That is, the capacity of a channel $\channel_{S A\rightarrow B}$ with  CSI at the decoder is the same as that of 
 a channel $\mathcal{M}'_{A\rightarrow S,B}$ without  parameters, where
\begin{align}
\mathcal{M}'_{A\rightarrow S,B}(\rho_A)=\sum_{s\in\Sset} q(s) \kb{s} \otimes \channel_{A\rightarrow B}^{(s)}(\rho_A) \,.
\end{align}
Hence, 
\begin{align}
\opC_{D}(\channel)=\opC(\mathcal{M}')= \max_{|\phi\rangle_{AA'}} I(A;B,S)_\rho=\max_{|\phi\rangle_{AA'}} I(A;B|S)_\rho 
\end{align}
with $\rho_{SAB}$ as in (\ref{eq:rhoSAB}), where the last equality holds by the chain rule and since $I(A;S)_\rho=0$ given that the Alice is not aware of the channel parameter.

Now, suppose that both Alice and Bob are aware of the channel parameter sequence.
Then, as explained above, the channel parameter $S$ can be thought of as part of the channel output in this case. Thus,  the corollary below immediately follows from Theorem~\ref{theo:mainNC}.
Define 
\begin{align}
\label{eq:inCeaED}
\inC_{ED,\text{n-c}}(\channel)\triangleq \max
I(A;B|S)_\omega 
\end{align}
where the maximization is as in (\ref{eq:inCea}).  
\begin{corollary}
\label{coro:mainNCed}
The entanglement-assisted classical capacity of the random-parameter quantum channel $\channel_{S A\rightarrow B}$ with non-causal CSI at both the encoder and the decoder is given by 
\begin{align}
\opC_{ED,\text{n-c}}(\channel)=\inC_{ED,\text{n-c}}(\channel) 
\end{align}
and the entanglement-assisted quantum capacity is given by 
$
\opQ_{ED,\text{n-c}}(\channel)=\frac{1}{2}\inC_{ED,\text{n-c}}(\channel) 
$. 
\end{corollary}
Based on our result in Theorem~\ref{theo:mainC}, we observe that the same capacity formula if valid for causal CSI as well.
To show achievability, set $\Fset^{(s)}$ to be clean, \ie $\Fset^{(s)}(\rho)=\rho$ for $s\in\Sset$. 
The converse part follows from Corollary~\ref{coro:mainNCed}, since the capacity with non-causal CSI is always an upper bound on the capacity with causal CSI.

\subsection{Discussion}
We give a few remarks on the results above. 
There is  clear similarity between the capacity formulas (\ref{eq:Cgp}) and (\ref{eq:inCea}) given non-causal CSI.
In particular, it can be seen that  the classical variables $U$ and  $X$ in  (\ref{eq:Cgp}) are replaced by the quantum systems 
$A$ and $A'$ in  (\ref{eq:inCea}), respectively. 
For the classical formula (\ref{eq:Cgp}), as shown in \cite{GelfandPinsker:80p,HeegardElGamal:83p}, the maximization  can be restricted to distributions $p_{U,X|S}=p_{U|S}p_{X|U,S}$ such that $p_{X|U,S}$ is a $0$-$1$ probability law, based on simple convexity arguments. The property stated in Lemma~\ref{lemm:pureCea} can thus be viewed as the quantum counterpart.

As for causal CSI, we observe that as in Shannon's classical proof for a classical channel with causal CSI \cite{Shannon:58p} 
\cite[Section 3.1]{KeshetSteinbergMerhav:07n}, our communication scheme can be interpreted as coding for a virtual channel $\Mset$, where the auxiliary plays the role of the channel input. Another similar trait is that at time $i$, the encoder applies a mapping that depends on the present $s_i$, while ignoring the sequence of past parameters, $s_1,\ldots,s_{i-1}$. In the classical setting, the mapping is the Shannon strategy $T(s_i)$, while in the quantum setting, it is the quantum channel $\Fset^{(s_i)}_{K\rightarrow A}$.

The classical capacity formula (\ref{eq:CClcaual}) for a classical channel with causal CSI can also be expressed as in (\ref{eq:Cgp}), constrained such that $U$ and $S$ are statistically independent \cite{Jafar:06p,KeshetSteinbergMerhav:07n}, and
the direct part can be proved by modifying the proof for non-causal CSI accordingly \cite[Section 7.6.3]{ElGamalKim:11b}.
In analogy, for a quantum channel, the classical variable $U$ is replaced by the quantum system $K$ in (\ref{eq:inCeaCausal}), where $K$ and $S$ are in a product state. 
Nonetheless, we observe that in the analysis, the causality requirement also dictates that Alice applies the encoding operations in  a different order compared to that of our  coding scheme with non-causal CSI (see Remark~\ref{rem:CreversedEop}).

\section*{Acknowledgment}
We gratefully thank Mark M. Wilde (Louisiana State University) for raising our attention to previous work by Dupuis \cite{Dupuis:08a,Dupuis:09c}.

The work was supported by the German Federal Ministry of Education and Research
 (Minerva Stiftung) and the Viterbi scholarship of the Technion.

\begin{appendices}

\section{Proof of Lemma~\ref{lemm:pureCeaCausal}}
\label{app:pureCeaCausal}
Fix the quantum state $\theta_{KA'}$ and channels  $\Fset^{(s)}_{K\rightarrow A}$, $s\in\Sset$, such that  
\begin{align}
\inC_{E,\text{caus}}(\channel)= I(K;B)_\omega 
\end{align}
and consider the spectral decomposition,
\begin{align}
\theta_{KA'}=\sum_{x\in\Xset}\sum_{z\in\Zset} p_{X,Z}(x,z) \kb{ x } \otimes \kb{ z } 
\end{align}
where $P_{X,Z}(x,z)$ is a  probability distribution, while $\{ |x\rangle \}_{x\in\Xset}$ and $\{ |z\rangle \}_{z\in\Zset}$ are orthonormal bases of the Hilbert spaces $\Hset_{K}$ and $\Hset_{A'}$, respectively.

To show that maximizing over pure states is sufficient, we  perform purification of the state $\theta_{KA'}$.
Specifically, define the pure state
\begin{align}
|\xi_{KJA'} \rangle= \sum_{x\in\Xset}\sum_{z\in\Zset} \sqrt{p_{X,Z}(x,z)}    |x\rangle \otimes 
|\psi_x\rangle \otimes |z\rangle   
\label{eq:AKJsC}
\end{align}
where $J$ is a reference system and $|\psi_x\rangle$ are orthonormal vectors in $\Hset_J$. Observe that $|\xi_{KJA'} \rangle$ is a purification of the mixed state $\theta_{KA'}$, namely, $\theta_{KA'}= \trace_J( \kb{ |\xi_{KJA'} \rangle } )$.
Defining $\widetilde{\Fset}^{(s)}_{KJ\rightarrow A}=(\Fset^{(s)}_{K\rightarrow A} \otimes\identity)$ and $\widetilde{K}=(K,J)$, we have that $\inC_{E,\text{caus}}(\channel)\geq I(\widetilde{K};B)$ by the definition  in (\ref{eq:inCeaCausal}). 
Yet, by the chain rule for the quantum mutual information \cite[Theorem 11.7.1]{Wilde:17b}, 
$\inC_{E,\text{caus}}(\channel)=$ $I(K;B)\leq$ $I(K,J;B)=$ $I(\widetilde{K};B)$. Hence, $\inC_{E,\text{caus}}(\channel)=$  $I(\widetilde{K};B)$.
Thereby, $\theta_{KA'}$ can be replaced by the pure state $|\xi_{\widetilde{K} A'} \rangle$.
\qed

\section{Proof of Theorem~\ref{theo:mainC}}
\label{app:mainC}

\subsection{Achievability Proof}
We show that for every $\eps_0,\delta_0>0$, there exists a $(2^{nR},n,\eps_0)$ code for the random-parameter quantum channel $\channel_{SA\rightarrow B}$ with causal CSI, provided that $R<\inC_{\text{caus}}(\channel)-\delta_0$. 
Based on Lemma~\ref{lemm:pureCeaCausal}, it suffices to consider a pure entangled state. 
Hence, let $|\xi_{KB} \rangle$ be a  pure entangled state, and $\Fset^{(s)}_{K\rightarrow A}(\rho_K)$, $s\in\Sset$, be a set of isometric channels.  Suppose that Alice and Bob share the joint state $|\xi_{KB} \rangle^{\otimes n}$.
Define the channel $\Mset_{K\rightarrow B'}$ by
\begin{align}
\Mset(\rho_K)=& \sum_{s\in\Sset}q(s)\channel^{(s)}\left( \Fset^{(s)}(\rho_K) \right)
\label{eq:ABphisC}
\end{align}
 and consider 
 the Schmidt decomposition of the state, 
\begin{align}
|\xi_{K,B}\rangle=\sum_{x\in\Xset} \sqrt{p_{X}(x)} |x\rangle \otimes |\psi_{x}\rangle
\end{align}
where $p_{X}$ is a probability distribution, $\{|x\rangle \}$ is an orthonormal basis of $\Hset_A$, and $|\psi_{x}\rangle$ are orthonormal vectors in $\Hset_B$. 

The code construction, encoding and decoding procedures are described below.

\subsubsection{Code Construction}
\begin{enumerate}[(i)]
\item
Select $2^{nR}$ independent sequences $x^n(m)$
at random, each according to $\prod_{i=1}^n p_X(x_i)$. 

\item
Quantum Operators: 
Consider the Heisenberg-Weyl operators $\{ \Sigma(a,b)= X(a)Z(b) \}$ of dimension $D$, given by
\begin{align}
X(a)=& \sum_{j=0}^{D-1} |a\oplus j\rangle \langle j |\\
Z(b)=& \sum_{j=0}^{D-1} e^{2\pi ibj/D} |j\rangle \langle j|
\end{align}
for $a,b\in \{ 0,1,\ldots, D-1  \}$, where 
$a\oplus j= (a+j) \mod D \;$ and
$i=\sqrt{-1}$.
For every  type class $\Tset_n(t)$ in $\Xset^n$, define the operators 
\begin{align}
V_{t}(a_t,b_t,c_t)= (-1)^{c_t} \Sigma(a_t,b_t) \,,\; a_t,b_t\in \{ 0,1,\ldots, D_t-1  \} \,,\; c_t\in\{0,1\} 
\label{eq:VtC}
\end{align}
where $D_t=|\Tset_n(t)|$ is the size of type class of $t$. Define the operator
\begin{align}
U(\gamma)=\bigoplus_t  \, V_{t}(a_t,b_t,c_t) 
\label{eq:UgammaC}
\end{align}
with $\gamma=\left( (a_t,b_t,c_t)_t \right)$, and let $\Gamma$ denote the set of all possible vectors $\gamma$. Then, choose 
$2^{nR}$ vectors $\gamma(m)$, $m\in [1:2^{nR}]$, uniformly at random.
\end{enumerate}

\begin{figure}
\begin{center}
\includegraphics[scale=0.7,trim={1cm 6.5cm 0 6cm},clip]{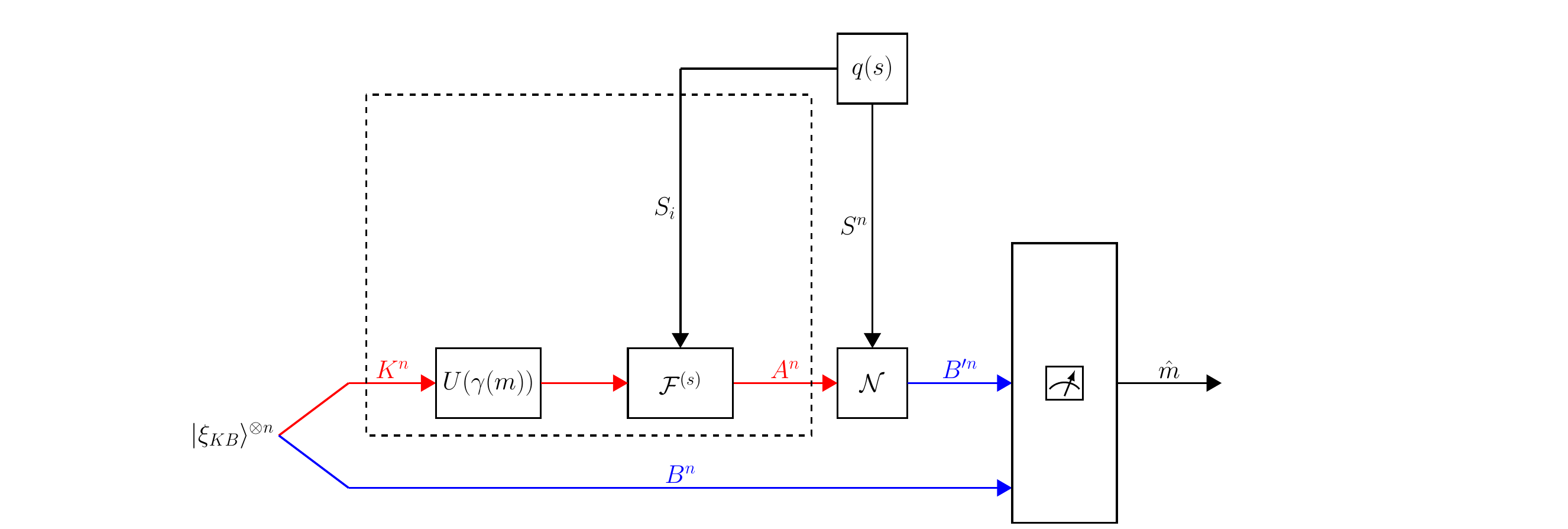} 
\end{center}
\caption{
Coding scheme with causal CSI at the encoder, using generalized super-dense coding for the virtual channel $\Mset_{K\rightarrow B'}$. The quantum systems of Alice and Bob are marked in red and blue, respectively. The blocks inside the dashed-line rectangle correspond to Alice's operations.
}
\label{fig:EAsiCodeDirC}
\end{figure}
\subsubsection{Encoding and Decoding}
The coding scheme is depicted in Figure~\ref{fig:EAsiCodeDirC}.
To send a message $m\in [1:2^{nR}]$, given a parameter sequence $s^n\in\Sset^n$, Alice performs the following.
\begin{enumerate}[(i)]

\item
 Apply the operator $U(\gamma(m))$ to $|\xi_{KB} \rangle^{\otimes n}$, which yields 
\begin{align}
|\varphi_{K^n B^n}^{m}\rangle\equiv (  U(\gamma(m)  \otimes\identity) |\xi_{KB} \rangle^{\otimes n} \,.
\end{align}

\item
Then, at time $i\in [1:n]$, apply the channel  $(\Fset^{(s_i)}\otimes\identity)$ to $|\varphi_{K_i B_i}^{m}\rangle$,  
and send the system $A_i$ through the channel.
\end{enumerate}

Bob receives the systems $B'^n$ at state $\omega_{B'^n B^n}$ and decodes the message by applying a POVM $\{ \Lambda_m \}_{m\in [1:2^{nR}]}$, which will be specified later. 

\subsubsection{Code Properties}
First, we write the entangled states as a combination of maximally entangled states over the typical subspaces, and then we can use the following useful identities.
For a maximally entangled state $
| \Phi_{AB} \rangle = \frac{1}{\sqrt{D}} \sum_{j=0}^{D-1} |j\rangle_A \otimes |j\rangle_B 
$,
\begin{align}
\trace_B \left( \kb{ \Phi_{AB} }  \right) =\pi_A
\end{align}
where  $\pi_A=\frac{1}{D}\sum_{x\in\Xset} \kb{x}$ is the maximally mixed state.
Furthermore, for every state $\rho$ of the system $A$, 
\begin{align}
\frac{1}{D^2}\sum_{a=0}^{D-1} \sum_{b=0}^{D-1} \Sigma(a,b)\, \rho \,\Sigma^{\dagger}(a,b) 
=\pi_A 
\label{eq:piIdenC}
\end{align}
(see \eg \cite{BennettShorSmolin:99p}  \cite[Exercise 4.7.6]{Wilde:17b})).
 Another useful identity is the ``ricochet property" \cite[Eq. (17)]{HsiehDevetakWinter:08p},
\begin{align}
(U\otimes\identity)|\Phi_{AB}\rangle=(\identity\otimes U^T)|\Phi_{AB}\rangle \,.
\label{eq:ricochetC}
\end{align}

Now, 
\begin{align}
|\xi_{K, B} \rangle^{\otimes n}
=& \sum_{x^n\in\Xset^n} \sqrt{ p_{X^n}(x^n) }  |x^n\rangle \otimes |\psi_{x^n} \rangle
\end{align}
where $p_{X^n}(x^n)=\prod_{i=1}^n p_{X}(x_i)$ and $|\psi_{x^n} \rangle=|\psi_{x_1} \rangle\otimes|\psi_{x_2} \rangle\otimes\cdots\otimes|\psi_{x_n} \rangle$. As the space $\Xset^n$ can be partitioned into  type classes, we may write 
\begin{align}
|\xi_{K, B} \rangle^{\otimes n}
=& \sum_{t\in\pSpace_n(\Xset)} \sum_{x^n\in \Tset_n(\hP)} \sqrt{ p_{X^n}(x^n) }  |x^n\rangle \otimes |\psi_{x^n} \rangle
\nonumber\\
=&  \sum_{t\in\pSpace_n(\Xset)} \sqrt{ p_{X^n}(x_t^n) } \sum_{x^n\in \Tset_n(t)}   |x^n\rangle \otimes |\psi_{x^n} \rangle
\end{align}
where $x_t^n$ is any sequence in the  type class $\Tset_n(t)$. Therefore, we have that 
\begin{align}
|\xi_{K, B} \rangle^{\otimes n}
=&  \sum_{t\in\pSpace_n(\Xset)} \sqrt{ P(t)}   |\Phi_t \rangle \,,
\label{eq:phiKBsnTC}
\end{align}
where
\begin{align}
& P(t)=d_t\cdot p_{X^n}(x_t^n)  \,, \;\text{with $d_t\equiv |\Tset_n(t)|$}
 \nonumber\\ 
& |\Phi_t \rangle=\frac{1}{\sqrt{d_t}} \sum_{x^n\in \Tset_n(t)}   |x^n\rangle \otimes |\psi_{x^n} \rangle
\end{align}
We note that $P(t)$ is the  probability of the type $\Tset_n(t)$ for a classical random sequence
$X^n\sim p_{X^n}$.

Now, Alice applies the operator $U(\gamma(m))$ to the entangled states.
Since the state $|\Phi_t \rangle$ is  maximally entangled, we have by the ``ricochet property" 
(\ref{eq:ricochetC}) that
\begin{align}
|\varphi_{K^n B^n}^{m}\rangle\equiv (U(\gamma(m))\otimes \identity)
|\xi_{K, B} \rangle^{\otimes n} =
(\identity\otimes U^T(\gamma(m)))|\xi_{K, B} \rangle^{\otimes n} \,.
\label{eq:RicoConsC}
\end{align}
That is, Alice's unitary operations can be reflected and treated as if performed by Bob. Then, Alice applies the channels 
$\Fset^{(s_i)}$ to her share of $|\varphi_{K_i B_i}^{m}\rangle$.

Subsequently, Bob  receives the systems $B'^n$ at state
\begin{align}
\rho^{\gamma(m)}_{ B'^n, B^n}=
&
\sum_{s^n\in\Sset^n} q^n(s^n)   (\channel^{(s^n)}\otimes\identity) (\Fset^{(s^n)}\otimes\identity)  \left( 
\left( \kb{ \varphi_{K^n B^n}^{m }}
\right)\right)
\label{eq:rhoG1C}
\\
=&
(\Mset^{\otimes n}\otimes\identity) \left( 
(\identity\otimes U^T(\gamma(m))) (\kb{ \xi_{K B} })^{\otimes n} (\identity\otimes U^*(\gamma(m)))
\right)
\label{eq:rhoG1cC}
\end{align}
where the last line is due to  (\ref{eq:RicoConsC}). 
Since a quantum channel is a linear map,  the above can be written as
\begin{align}
\rho^{\gamma}_{B'^n, B^n}
=& 
(\identity\otimes U^T(\gamma))  \left[ \left( (\Mset\otimes\identity)  (\kb{ \xi_{KB} }) \right)^{\otimes n} \right]
(\identity\otimes U^*(\gamma))
\nonumber\\
=& 
(\identity\otimes U^T(\gamma))  \omega_{B'B}^{\otimes n}
(\identity\otimes U^*(\gamma))
\label{eq:rhoG2C}
\end{align}
where we have defined
\begin{align}
&\omega_{B'B}= ( \Mset\otimes \identity)(\kb{\xi_{KB}}) \,.
\label{eq:oBpBC}
\end{align}

\subsubsection{Packing Lemma Requirements}
Next, we use the quantum packing lemma. Consider the ensemble 
$\left\{ p(\gamma)=\frac{1}{|\Gamma|},  \rho^{\gamma}_{B'^n, B^n} \right\}$, for which the expected density operator is 
\begin{align}
\sigma_{B'^n,B^n}=\frac{1}{|\Gamma|}\sum_{\gamma\in\Gamma} \rho^{\gamma}_{B'^n, B^n} \,.
\label{eq:SigmaBpBC}
\end{align}
Define the code projector and the codeword projectors by
\begin{align}
 \Pi \equiv& \Pi^{\delta}(\omega_{B'})\otimes \Pi^{\delta}(\omega_{B})
\label{eq:CodeProjDirC}
\\
 \Pi_\gamma \equiv& (\identity\otimes U^T(\gamma)) \Pi^{\delta}(\omega_{B'B}) (\identity\otimes U^*(\gamma)) \,,\; \text{for 
$\gamma\in\Gamma$}
\end{align}
where $\Pi^{\delta}(\omega_{B'B})$, $\Pi^{\delta}(\omega_{B'})$ and $\Pi^{\delta}(\omega_{B})$ are the projectors onto the $\delta$-typical subspaces associated with the states $\omega_{B'B}$, $\omega_{B'}=\trace_B(\omega_{B'B})$ and
$\omega_{B}=\trace_{B'}(\omega_{B'B})$, respectively (see (\ref{eq:oBpBC})).
Now, we verify that the assumptions of Lemma~\ref{lemm:Qpacking} hold with respect to the ensemble and the projectors above.

First, we show that $\trace(\Pi \rho^{\gamma}_{B'^n, B^n})\geq\, 1-\alpha$, where $\alpha>0$ is arbitrarilly small. Defining 
$\check{P}=\identity-P$, we have that
\begin{align}
\Pi= (\identity-\check{\Pi}^{\delta}(\rho_{B'}))\otimes (\identity- \check{\Pi}^{\delta}(\rho_{B}) )
\succeq (\identity\otimes\identity)-(\check{\Pi}^{\delta}(\rho_{B'})\times\identity)-(\identity\otimes\check{\Pi}^{\delta}(\rho_{B}))
\end{align}
hence,
\begin{align}
\trace(\Pi \rho^{\gamma}_{B'^n, B^n})\geq& 1
-\trace\left( (\check{\Pi}^{\delta}(\rho_{B'})\otimes\identity) \rho^{\gamma}_{B'^n, B^n} \right)
-\trace\left( (\identity\otimes\check{\Pi}^{\delta}(\rho_{B})) \rho^{\gamma}_{B'^n, B^n} \right)
\nonumber\\
=& 1
-\trace\left( \check{\Pi}^{\delta}(\rho_{B'}) \rho^{\gamma}_{B'^n} \right)
-\trace\left( \check{\Pi}^{\delta}(\rho_{B}) \rho^{\gamma}_{ B^n} \right) \,.
\label{eq:ReqPackDir1C}
\end{align}
The first trace term in the RHS of (\ref{eq:ReqPackDir1C}) equals $\trace\left( \check{\Pi}^{\delta}(\rho_{B'}) \omega_{B'}^{\otimes n} \right)$ by 
(\ref{eq:rhoG2C}), and the last term equals $\trace\left( \check{\Pi}^{\delta}(\rho_{B})  \omega_{B}^{\otimes n}  \right)$ by (\ref{eq:rhoG1C})
and (\ref{eq:oBpBC}). 
Therefore, we have by (\ref{eq:UnitT}) that 
\begin{align}
\trace(\Pi \rho^{\gamma}_{B'^n, B^n})\geq& 
1
-\trace\left( \check{\Pi}^{\delta}(\rho_{B'}) \omega_{B'}^{\otimes n} \right)
-\trace\left( \check{\Pi}^{\delta}(\rho_{B}))  \omega_{B}^{\otimes n}  \right)
\nonumber\\
\geq& 1-2\eps \,.
\end{align}
Similarly, the second requirement of the packing lemma  holds since
\begin{align}
\trace(\Pi_\gamma \rho^{\gamma}_{B'^n, B^n})=&\trace\left[ (\identity\otimes U^T(\gamma)) \Pi^{\delta}(\omega_{B'B}) (\identity\otimes U^*(\gamma))
(\identity\otimes U^T(\gamma))  \omega_{B'B}^{\otimes n}
(\identity\otimes U^*(\gamma))
  \right]
	\nonumber\\
	=&\trace( \Pi^{\delta}(\omega_{B'B})\omega_{B'B}^{\otimes n}) \geq 1-\eps 
\end{align}
where the second equality follows from the cyclicity of the trace and the fact that $U^* U^T=(U U^\dagger)^*=\identity$ for a unitary operator, and the last inequality is due to (\ref{eq:UnitT}).

Moving to the third requirement in Lemma~\ref{lemm:Qpacking},
\begin{align}
\trace(\Pi_\gamma)=\trace\left( (\identity\otimes U^T(\gamma)) \Pi^{\delta}(\omega_{B'B}) (\identity\otimes U^*(\gamma)) \right)
=\trace( \Pi^{\delta}(\omega_{B'B}) ) \leq 2^{n(H(\omega_{B'B})+c\delta)}
\end{align}
where the second equality holds by cyclicity of the trace and the last inequality is due to (\ref{eq:Pidim}).
It is left to verify that the last requirement of the packing lemma holds, \ie
$\Pi \sigma_{B'^n,B^n} \Pi \preceq\, 2^{-n(H(\sigma_{B'})+H(\sigma_B)-\alpha)} \Pi$. To this end, observe that by (\ref{eq:rhoG1cC}) and 
(\ref{eq:SigmaBpBC}),
\begin{align}
\sigma_{B'^n,B^n}=  (\Mset^{\otimes n} \otimes\identity) \tau_{K^n,B^n}
\label{eq:sigmaxiBpC}
\end{align}
where we have defined
\begin{align}
\tau_{K^n,B^n}\equiv \frac{1}{|\Gamma|} \sum_{\gamma\in\Gamma} (\identity\otimes U^T(\gamma))  (\kb{ \xi_{K B} })^{\otimes n} 
(\identity\otimes U^*(\gamma)) \,.
\label{eq:xisnC}
\end{align}
Then, by (\ref{eq:RicoConsC}) along with (\ref{eq:VtC})-(\ref{eq:UgammaC}),
\begin{align}
\tau_{K^n,B^n}
=&  \frac{1}{|\Gamma|} \sum_{\gamma\in\Gamma} (\identity\otimes U^T(\gamma)) \left(   \sum_{t} \sqrt{ P(t|s^n)}
   |\Phi_t \rangle  \right)
\left(   \sum_{t'} \sqrt{ P(t')}   \langle\Phi_{t'} | \right)  
(\identity\otimes U^*(\gamma)) 
\nonumber\\
=&  \frac{1}{|\Gamma|} \sum_{\gamma\in\Gamma} 
\left(   \sum_{t} \sqrt{ P(t)} (-1)^{c_t(\gamma)}  (\identity\otimes \Sigma^T_{a_t(\gamma),b_t(\gamma)})
   |\Phi_t \rangle  \right)
	\nonumber\\
	&\quad
\left(   \sum_{t'} \sqrt{ P(t')}  (-1)^{c_{t'}(\gamma)}   \langle\Phi_{t'} |  (\identity\otimes \Sigma^*_{a_{t'}(\gamma),b_{t'}(\gamma)}) \right)
\,.
\label{eq:xisn1C}
\end{align}
For $t'=t$, the expression above becomes
\begin{align}
 &\frac{1}{|\Gamma|} \sum_{\gamma\in\Gamma}   \sum_{t}  P(t)
 (\identity\otimes \Sigma^T_{a_t(\gamma),b_t(\gamma)})
  \kb{ \Phi_t }   (\identity\otimes  \Sigma^*_{a_{t}(\gamma),b_{t}(\gamma)}) 
	\nonumber\\
	=& \sum_{t}  P(t|s^n) \left[ \frac{1}{D_t^2} \sum_{a_t,b_t}  (\identity\otimes \Sigma^T_{a_t,b_t})
  \kb{ \Phi_t }   (\identity\otimes  \Sigma^*_{a_{t},b_{t}}) \right]
	= \sum_{t}  P(t) \pi^t_{K^n} \times \pi^t_{B^n}
	\label{eq:xisn2C}
\end{align}
with
\begin{align}
\pi^t_{K^n}\equiv\frac{\Pi_{K^n}(t)}{\trace(\Pi_{K^n}(t))} \,,\;
\pi^t_{B^n}\equiv \frac{\Pi_{B^n}(t)}{\trace(\Pi_{B^n}(t))}
\label{eq:pitKBnC}
\end{align}
where $\Pi_{K^n}(t)$ is the projector of type $t$ as defined in (\ref{eq:ProjType}).
 The last equality in (\ref{eq:xisn2C}) follows from (\ref{eq:piIdenC}). On the other hand, for $t'\neq t$, 
\begin{align}
  &\frac{1}{|\Gamma|} \sum_{\gamma\in\Gamma} 
   \sum_{t} \sum_{t'\neq t} \sqrt{ P(t)  P(t')} 
	\frac{1}{4D_t^2 D_{t'}^2} \sum_{c_t,c_{t'}\in \{0,1\}}
	(-1)^{c_t+c_{t'}} \sum_{a_t,a_{t'},b_t,b_{t'}}  (\identity\otimes \Sigma^T_{a_t,b_t})
   |\Phi_t \rangle  
    \langle\Phi_{t'} |  (\identity\otimes \Sigma^*_{a_{t'},b_{t'}}) 
		= 0
\,.
\label{eq:xisn3C}
\end{align}
We deduce from (\ref{eq:xisn1C})-(\ref{eq:xisn3C}) that $
\tau_{K^n,B^n}= \sum_{t}  P(t) \pi^t_{K^n} \otimes \pi^t_{B^n}
$. Plugging this into (\ref{eq:sigmaxiBpC}) yields
\begin{align}
\sigma_{B'^n,B^n}=  \sum_{t}  P(t) \Mset^{\otimes n}( \pi^t_{K^n})\otimes  \pi^t_{B^n} \,.
\end{align}

Now, we use the formula above in order to show that the last requirement in Lemma~\ref{lemm:Qpacking} holds. Consider that
\begin{align}
\Pi \sigma_{B'^n,B^n} \Pi =& (\Pi^{\delta}(\omega_{B'})\otimes \Pi^{\delta}(\omega_{B})) \sigma_{B'^n,B^n}
 (\Pi^{\delta}(\omega_{B'})\otimes \Pi^{\delta}(\omega_{B}))
\nonumber\\
=&   \sum_{t}  P(t) 
\big( \Pi^{\delta}(\omega_{B'})\Mset^{\otimes n}( \pi^t_{A^n}) \Pi^{\delta}(\omega_{B'}) \big)\otimes
\big( \Pi^{\delta}(\omega_{B}) \pi^t_{B^n} \Pi^{\delta}(\omega_{B}) \big) \,.
\end{align}
Using (\ref{eq:pitKBnC}), this can be bounded by
\begin{align}
\Pi \sigma_{B'^n,B^n} \Pi 
=&   \sum_{t}  P(t) 
\big( \Pi^{\delta}(\omega_{B'})\Mset^{\otimes n}( \pi^t_{K^n}) \Pi^{\delta}(\omega_{B'}) \big)\otimes
\big( \Pi^{\delta}(\omega_{B}) \frac{\Pi_{B^n}(t)}{\trace(\Pi_{B^n}(t))} \Pi^{\delta}(\omega_{B}) \big)
\nonumber\\
\preceq&
2^{-n(H(\omega_B)+\eps_1)}   \sum_{t}  P(t) 
\big( \Pi^{\delta}(\omega_{B'})\Mset^{\otimes n}( \pi^t_{K^n}) \Pi^{\delta}(\omega_{B'}) \big)\otimes
 \Pi^{\delta}(\omega_{B}) 
\end{align}
with arbitrarily small $\eps_1>0$, following (\ref{eq:Tsize}) and the fact that
$\Pi^{\delta}(\omega_{B})\Pi_{B^n}(t)\Pi^{\delta}(\omega_{B}) \preceq \Pi^{\delta}(\omega_{B})$.
By linearity, this can also be written as
\begin{align}
\Pi \sigma_{B'^n,B^n} \Pi 
\preceq&
2^{-n(H(\omega_B)+\eps_1)} 
 \Pi^{\delta}(\omega_{B'}) \left[ \Mset^{\otimes n}\left( \sum_{t}  P(t) \pi^t_{K^n} \right) \right]
 \Pi^{\delta}(\omega_{B'})  \otimes
 \Pi^{\delta}(\omega_{B}) 
\nonumber\\
=&
2^{-n(H(\omega_B)+\eps_1)} 
 \Pi^{\delta}(\omega_{B'}) \left[ \Mset^{\otimes n}\left( \omega_{K^n} \right) \right]
 \Pi^{\delta}(\omega_{B'})  \otimes
 \Pi^{\delta}(\omega_{B}) 
\end{align}
(see (\ref{eq:phiKBsnTC})). Since the expression in the square brackets equals $\omega_{B'}^{\otimes n}$ (see (\ref{eq:oBpBC})), we have by 
 (\ref{eq:rhonProjIneq})  that 
\begin{align}
\Pi \sigma_{B'^n,B^n} \Pi \preceq 2^{-n(H(\omega_B')+H(\omega_B)+\eps_1+\eps_2)}  \Pi^{\delta}(\omega_{B'})\otimes  \Pi^{\delta}(\omega_{B})
=2^{-n(H(\omega_B')+H(\omega_B)+\eps_1+\eps_2)} \Pi 
\end{align}
with arbitrarily small $\eps_2>0$,
where the last equality follows from the definition of $\Pi$ in (\ref{eq:CodeProjDirC}). It follows that all of the requirements of the packing lemma are satisfied. 

Hence, by Lemma~\ref{lemm:Qpacking}, there exist deterministic vectors $\gamma(m)$, $m\in [1:2^{nR}]$, and  a POVM 
$\{ \Lambda_{m} \}_{m\in [1:2^{n\tR}]}$ such that 
\begin{align}
  \trace\left( \Lambda_{m} \rho^{\gamma(m)}_{ B'^n, B^n} \right)  \geq 1-2^{-n[ I(B';B)_\omega-R-\eps']}
	\label{eq:packDirIneqC}
\end{align}
for all 
$m\in [1:2^{nR}]$, where $\eps'$ is arbitrarily small. 
That is,
the probability of error is bounded by $2^{-n[ I(B';B)_\omega-R-\eps_n(\alpha)]}$, which tends to zero if
\begin{align}
R< I(B';B)_\omega-\eps' \,.
\end{align}

Now, consider the systems $S,K_1,A_1,A_1',B_1$ at state
\begin{align}
&\omega^s_{  A_1 A_1' }= (\Fset_1^{(s)}\otimes\identity  )(\kb{ \xi_{K_1,A_1'}  })  \label{eq:oSKB0C} \\
&\omega_{A_1 S  A_1' }=\sum_{s\in\Sset} q(s) \kb{ s } \otimes  \omega^s_{  A_1 A_1' } \\
&\omega_{ A_1 B_1}= (\identity\otimes \channel)(\omega_{ A_1 S A_1'})=(\identity\otimes \Mset)(\kb{ \xi_{K_1,A_1'}  }) \,.
\label{eq:oKAB10C}
\end{align}
Observe that this is the same relation as in (\ref{eq:oBpBC}) where  $A_1'$,  $K_1$ and $B_1$
take place with  $A$,   $B$, and $B'$, respectively, where $\Fset_1^{(s)}$ is defined with Kraus operators $F_{1,j}^{(s)}\equiv (F_j^{(s)})^T$ for $s\in\Sset$, due to (\ref{eq:phiKBsnTC}) and the ``ricochet property" (\ref{eq:ricochetC}).
Thus, the probability of error tends to zero as $n\rightarrow\infty$ provided that  $R<I(K_1;B_1)_\omega-\eps'$.
This completes the proof of the direct part.

\subsection{Converse Proof}
Consider the converse part. Suppose that Alice and Bob are trying to distribute randomness. An upper bound on the rate at which Alice can distribute randomness to Bob also serves as an upper bound on the rate at which they can communicate. In this task, Alice and Bob share an entangled state $\Psi^{\otimes n}_{T_A T_B}$. Alice first prepares the maximally corrleated state
\begin{align}
\overline{\Phi}_{MM'} \equiv \frac{1}{2^{nR}}\sum_{m=1}^{2^{nR}} \kb{ m } \otimes \kb{ \phi_m } \,.
\end{align}
locally. We note that since $M$ and $M'$ are classical, they can be copied. 
 
 Then, at time $i\in [1:n]$, Alice applies an encoding channel $\Eset^{s^i}_{M'T_{Ai}\rightarrow A_i'}$ to the classical system $M'$ and her share $T_{Ai}$ of the entangled state $\Psi^{\otimes n}_{T_A T_B}$. 
The resulting state is $\omega_{S^i M A'_i T_{Bi}}=\sum_{s^i\in\Sset^i} q^i(s^i) \kb{ s^i }\otimes 
\rho^{s^i}_{ M A'_i T_{Bi}}$, with
\begin{align}
\rho^{s^i}_{ M  A'_i T_{Bi} }\equiv (\identity\otimes\Eset^{s^i}\otimes\identity)( \overline{\Phi}_{MM'}\otimes \Psi_{T_{Ai} T_{Bi}} ) \,,
\label{eq:siCausConv1C}
\end{align}
for $i\in [1:n]$.
After Alice sends the systems $A'^n$ through the channel, Bob receives the systems $B^n$ at state
$\omega_{S^n M B^n T_B^n}=\sum_{s^n\in\Sset^n} q^n(s^n) \kb{ s^n }\otimes 
\rho^{s_1}_{ M B_1 T_{B1}}\otimes \rho^{s_1,s_2}_{ M B_2 T_{B2}}\otimes\cdots\otimes \rho^{s^n}_{ M B_n T_{Bn}}$, with
\begin{align}
\rho^{s^i}_{M B_i T_{Bi}}\equiv (\identity\otimes\channel^{(s_i)}\otimes\identity) (\rho^{s^i}_{ M  A'_i T_{Bi} }) \,,
\end{align}
for $i\in [1:n]$.
Then, Bob performs a decoding channel $\Dset_{B^n T_B^n\rightarrow \hM}$, producing $\omega_{S^n M \hM}'=\sum_{s^n\in\Sset^n} q^n(s^n) \kb{ s^n }\otimes 
\rho^{s^n}_{ M \hM}$ with
\begin{align}
\rho^{s^n}_{ M \hM}\equiv (\identity\otimes\Dset)\left(\bigotimes_{i=1}^n \rho^{s^i}_{M B_i T_{Bi}} \right)
\label{eq:DecConv1C}
\end{align}

Consider a sequence of codes $(\Eset_n^{s^i},\Psi_n,\Dset_n)$ for randomness distribution, such that
\begin{align}
\frac{1}{2} \norm{ \omega_{M\hM} -\overline{\Phi}_{MM'} }_1 \leq \alpha_n \,,
\end{align}
where $\omega_{M\hM}$ is the reduced density operator of $\omega_{S^n M \hM}$ and while $\alpha_n$ tends to zero as $n\rightarrow\infty$.
By the Alicki-Fannes-Winter inequality \cite{AlickiFannes:04p,Winter:16p} \cite[Theorem 11.10.3]{Wilde:17b}, this implies that
\begin{align}
|H(M|\hM)_\omega - H(M|M')_{\overline{\Phi}} |\leq n\eps_n
\label{eq:AFWC}
\end{align}
while $\eps_n$ tends to zero as $n\rightarrow\infty$.
Now, observe that $H(\overline{\Phi}_{M M'})=H(\overline{\Phi}_{M})=H(\overline{\Phi}_{ M'})=nR$, hence $I(M;\hM)_{\overline{\Phi}}=nR$.
Also,   $H(\omega_{M})=H(\overline{\Phi}_{ M})=nR$ implies that 
$I(M;M')_{\overline{\Phi}} - I(M;\hM)_{\omega}= H(M|\hM)_\omega - H(M|M')_{\overline{\Phi}}$. Therefore, by (\ref{eq:AFWC}),
\begin{align}
nR=&I(M;\hM)_{\overline{\Phi}} \nonumber\\
\leq& I(M;\hM)_{\omega}+n\eps_n \nonumber\\
\leq& I(M;T_B,B^n)_{\omega}+n\eps_n 
\label{eq:ConvIneq1C}
\end{align}
where the last line follows from (\ref{eq:DecConv1C}) and the quantum data processing inequality \cite[Theorem 11.5]{NielsenChuang:02b}.

As in the classical case, the chain rule for the quantum mutual information states that 
$I(A;B,C)_\sigma=I(A;B)_\sigma+I(A;C|B)_\sigma$ for all $\sigma_{ABC}$ (see \eg \cite[Property 11.7.1]{Wilde:17b}). Hence,
\begin{align}
nR \leq& I(T_B,M;B^n)_{\omega}+I(M;T_B)_{\omega}-I(T_B;B^n)_{\omega}+n\eps_n \nonumber\\
\leq& I(T_B,M;B^n)_{\omega}+I(M;T_B)_{\omega}+n\eps_n \nonumber\\
=& I(T_B,M;B^n)_{\omega}+n\eps_n 
\label{eq:ConvIneq2C}
\end{align}
where the equality holds since the systems $M$ and $T_B$ are in a product state. The chain rule further 
implies that 
\begin{align}
I(T_B,M;B^n)_{\omega}=&\sum_{i=1}^n I(T_{B},M;B_i| B^{i-1})_\omega 
\nonumber\\
\leq& \sum_{i=1}^n I(T_B,M,S^{i-1},A'^{i-1},B^{i-1};B_i)_\omega \nonumber\\
=&  \sum_{i=1}^n [ I(T_B,M,S^{i-1},A'^{i-1};B_i)_\omega + I(B^{i-1};B_i|T_B,M,S^{i-1},A^{i-1})_\omega ] \nonumber\\
=&  \sum_{i=1}^n  I(T_B,M,S^{i-1},A'^{i-1};B_i)_\omega
\label{eq:ConvIneq3C}
\end{align}
where the last line holds since the channel has a product form, \ie $\channel_{S^i A'^i\rightarrow B^i}= \channel^{\otimes i}=
\channel_{S^{i-1} A'^{i-1}\rightarrow B^{i-1}} \otimes \channel_{S_i A_i'\rightarrow B_i} $. 
Defining $K_i=(M,M',S^{i-1},A'^{i-1},T_A,T_B)$ and a quantum channel $\Fset^{(s_i)}_{K_i\rightarrow A_i}$, 
we have by (\ref{eq:ConvIneq2C}) and (\ref{eq:ConvIneq3C}) that
\begin{align}
R-\eps_n\leq \frac{1}{n}\sum_{i=1}^n  I(K_i;B_i)_\omega 
\leq \max_{\theta_{KA'}\,,\; \Fset^{(s)}_{K\rightarrow A}}  I(K;B)_\omega  
\,.
\end{align}
Observe that by (\ref{eq:siCausConv1C}), $K_i$ and $S_i$ are in a product state as required.
This concludes the proof of Theorem~\ref{theo:mainC}.
\qed 

\section{Proof of Lemma~\ref{lemm:pureCea}}
\label{app:pureCea}
Fix the quantum state $\theta_{KA'}$ and channels  $\Fset^{(s)}_{K\rightarrow A}$, $s\in\Sset$, such that  
\begin{align}
\inC_{E,\text{n-c}}(\channel)= I(A;B)_\omega-I(A;S)_\omega 
\end{align}
and consider the spectral decomposition,
\begin{align}
\theta_{KA'}=\sum_{x\in\Xset}\sum_{z\in\Zset} p_{X,Z}(x,z) \kb{ x } \otimes \kb{ z } 
\end{align}
where $P_{X,Z}(x,z)$ is a  probability distribution, while $\{ |x\rangle \}_{x\in\Xset}$ and $\{ |z\rangle \}_{z\in\Zset}$ are orthonormal bases of the Hilbert spaces $\Hset_{K}$ and $\Hset_{A'}$, respectively.
Also, for every $s\in\Sset$, consider the Kraus representation of each channel
\begin{align}
\Fset^{(s)}_{K\rightarrow A}(\rho_K)= \sum_{j} F_j^{(s)} \rho_A F_j^{(s)\,\dagger} 
\end{align}
with $\sum_j F_j^{(s)\,\dagger} F_j^{(s)}=\identity$ (see Subsection~\ref{subsec:Qchannel}).

First, we show that maximizing over pure states is sufficient.
To this end, we  perform purification of the state $\theta_{KA'}$.
Specifically, define the pure state
\begin{align}
|\xi_{KJA'} \rangle= \sum_{x\in\Xset}\sum_{z\in\Zset} \sqrt{p_{X,Z}(x,z)}    |x\rangle \otimes 
|\psi_x\rangle \otimes |z\rangle   
\label{eq:AKJs}
\end{align}
where $J$ is a reference system and $|\psi_x\rangle$ are orthonormal vectors in $\Hset_J$. Observe that $|\xi_{KJA'} \rangle$ is a purification of the mixed state $\theta_{KA'}$, namely, $\theta_{KA'}= \trace_J( \kb{ |\xi_{KJA'} \rangle } )$.
Defining $\widetilde{\Fset}^{(s)}_{KJ\rightarrow A}=(\Fset^{(s)}_{K\rightarrow A} \otimes\identity)$ and $\widetilde{K}=(K,J)$, we have that
\begin{align}
\omega_{AA'}= \widetilde{\Fset}^{(s)}_{\widetilde{K}\rightarrow A}(\kb{ \xi_{\widetilde{K} A'}  }) \,.
\end{align}
Then, observe that the mutual information difference
$[ I(A;B)_\omega-I(A;S)_\omega ] $ depends on the state $\theta_{KA'}$ and the channels $\Fset^{(s)}_{K\rightarrow A}$ only through $\omega_{AA'}$, and thus, $\theta_{KA'}$ can be replaced by the pure state $|\xi_{\widetilde{K} A'} \rangle$.

To show that maximizing over isometric channels is sufficient, we use an  isometric extension of the channels 
$\Fset_{K\rightarrow A}^{(s)}$, for $s\in\Sset$.
Define the isometric channels $\overline{\Fset}_{K\rightarrow AE}^{(s)}$ by
\begin{subequations}
\label{eq:FKAE}
\begin{align}
\overline{\Fset}^{(s)}_{K \rightarrow AE}(\rho_K)=\overline{F}^{(s)} \rho_K \overline{F}^{(s)\,\dagger} 
\end{align}
for all $\rho_K$, with
\begin{align}
\overline{F}^{(s)}=\sum_j F_j^{(s)} \otimes |j\rangle
\end{align}
\end{subequations}
where $E$ is a reference system  and $\{|j\rangle \}$ is an orthonormal basis of $\Hset_E$.
Observe that $\overline{\Fset}^{(s)}_{K \rightarrow AE}$ is an extension of the quantum channel $\Fset^{(s)}_{K\rightarrow A}$, namely,
$\trace_E\left( \overline{\Fset}^{(s)}_{K \rightarrow AE}(\rho_K) \right)=\Fset^{(s)}_{K\rightarrow A}(\rho_K)$ for every $\rho_K$.

Let 
\begin{align}
&\sigma^s_{AEA'}=  (\overline{\Fset}^{(s)}_{K\rightarrow AE} \otimes \identity   )(\theta_{KA'}) \\
&\sigma_{AESA'}=\sum_{s\in\Sset} q(s) \kb{ s } \otimes \sigma^s_{AEA'}  \\
&\sigma_{AEB}= (\identity\otimes\identity\otimes \channel)(\sigma_{AESA'}) \,.
\end{align}
Based on the definition in (\ref{eq:inCea}),
\begin{align}
\inC_{E,\text{n-c}}(\channel)\geq I(A,E;B)_\sigma-I(A,E;S)_\sigma \,.
\end{align}
On the other hand, by the quantum data processing theorem due to Schumacher and Nielsen \cite{SchumacherNielsen:96p}\cite[Theorem 11.9.4]{Wilde:17b},
\begin{align}
I(A;B)_\omega  \leq I(A,E;B)_\sigma \,.
\label{eq:KBl}
\end{align}
Furthermore, by (\ref{eq:FKAE}),  the systems $S$ and  $E$ are in a product state given $A$, hence
 $I(E;S|A)_\sigma=0$. Thus,
\begin{align}
I(A;S)_\omega =I(A;S)_\sigma = I(A;S)_\sigma+ I(E;S|A)_\sigma= I(A,E;S)_\sigma 
\label{eq:KSl}
\end{align}
where the last equality is due to the chain rule for the quantum mutual information \cite[Theorem 11.7.1]{Wilde:17b}.
Together, (\ref{eq:KBl}) and (\ref{eq:KSl}) imply that 
\begin{align}
\inC_{E,\text{n-c}}(\channel)\leq I(A,E;B)_\sigma-I(A,E;S)_\sigma \,.
\end{align}
It thus follows that the channel $\Fset_{K\rightarrow A}^{(s)}$ in (\ref{eq:inCea}) can be replaced by its isometric extension  
$\overline{\Fset}^{(s)}_{K \rightarrow A_0}$,  with $A_0=(A,E)$, for $s\in\Sset$. This completes the proof of the lemma.
\qed

\section{Proof of Theorem~\ref{theo:mainNC}}
\label{app:mainNC}

\subsection{Achievability Proof}
We show that for every $\eps_0,\delta_0>0$, there exists a $(2^{nR},n,\eps_0)$ code for the random-parameter quantum channel $\channel_{SA\rightarrow B}$ with non-causal CSI, provided that $R<\inC_{E,\text{n-c}}(\channel)-\delta_0$. 
Based on Lemma~\ref{lemm:pureCea}, it suffices to consider a pure entangled state and isometric channels.
Hence, let $|\xi_{AB} \rangle$ be a  pure entangled state, and $\Fset^{(s)}_{K\rightarrow A}(\rho_K)=F^{(s)}\rho_K F^{(s)\,\dagger}$, $s\in\Sset$, be a set of isometric channels.  Suppose that Alice and Bob share the joint state $|\xi_{AB} \rangle^{\otimes n}$.
Define
\begin{align}
|\varphi^s_{AB}\rangle=& (F^{(s)}\otimes\identity) |\xi_{AB} \rangle 
\label{eq:ABphis}
\end{align}
 and consider 
 the Schmidt decomposition of the state, 
\begin{align}
|\varphi_{A,B}^s\rangle=\sum_{x\in\Xset} \sqrt{p_{X|S}(x|s)} |x\rangle \otimes |\psi_{x,s}\rangle
\end{align}
where $p_{X|S}$ is a conditional probability distribution, $\{|x\rangle \}$ is an orthonormal basis of $\Hset_A$, and $|\psi_{x,s}\rangle$ are orthonormal vectors in $\Hset_B$. 
Observe that the quantum entropy of the system  $B$ is the same as the Shannon entropy of the classical random variable $X$, \ie
$
H(\omega_{SB})=H(S,X) 
$ and
$
H(\omega_B)=H(X) 
$. 
Thus,
\begin{align}
I(B;S)_{\varphi}=&I(X;S) 
\,.
\label{eq:IBS}
\end{align}

The code construction, encoding and decoding procedures are described below.

\subsubsection{Code Construction}
Encoding is performed in two stages, first classical compression of the parameter sequence $S^n$, and then, application of quantum operators depending on 
the result in the first stage. The code construction is specified below.
\begin{enumerate}[(i)]
\item
Classical Compression: 
Let $\tR>R$. We construct $2^{nR}$ sub-codebooks at random.
 For every message $m\in [1:2^{nR}]$,  choose $2^{n(\tR-R)}$ independent sequences $x^n(\ell)$
at random, each according to $\prod_{i=1}^n p_X(x_i)$. Then, we have the following  sub-codebooks, 
  \begin{align}
	\mathscr{B}(m)=\{  x^n(\ell) \,:\; \ell\in [(m-1)2^{n(\tR-R)}+1:m2^{n(\tR-R)}] \}  \,,\; \text{for $m\in [1:2^{nR}]$} \,.
	\end{align}

\item
Quantum Operators: 
Consider the Heisenberg-Weyl operators $\{ \Sigma(a,b)= X(a)Z(b) \}$ of dimension $D$, given by
\begin{align}
X(a)=& \sum_{j=0}^{D-1} |a\oplus j\rangle \langle j |\\
Z(b)=& \sum_{j=0}^{D-1} e^{2\pi ibj/D} |j\rangle \langle j|
\end{align}
for $a,b\in \{ 0,1,\ldots, D-1  \}$, where 
$a\oplus j= (a+j) \mod D \;$ and
$i=\sqrt{-1}$.
For every $s^n\in\Sset^n$ and every conditional type class $\Tset_n(t|s^n)$ in $\Xset^n$, define the operators 
\begin{align}
V_{t}(a_t,b_t,c_t)= (-1)^{c_t} \Sigma(a_t,b_t) \,,\; a_t,b_t\in \{ 0,1,\ldots, D_t-1  \} \,,\; c_t\in\{0,1\} 
\label{eq:Vt}
\end{align}
where $D_t=|\Tset_n(t|s^n)|$ is the size of type class associated with the conditional type $t$. Then, define the operator
\begin{align}
U(\gamma)=\bigoplus_t  \, V_{t}(a_t,b_t,c_t) 
\label{eq:Ugamma}
\end{align}
with $\gamma=\left( (a_t,b_t,c_t)_t \right)$. Let $\Gamma$ denote the set of all possible vectors $\gamma$. Then, choose 
$2^{n\tR}$ vectors $\gamma(\ell)$, $\ell\in [1:2^{n\tR}]$, uniformly at random.
\end{enumerate}

\begin{figure}
\begin{center}
\includegraphics[scale=0.4]{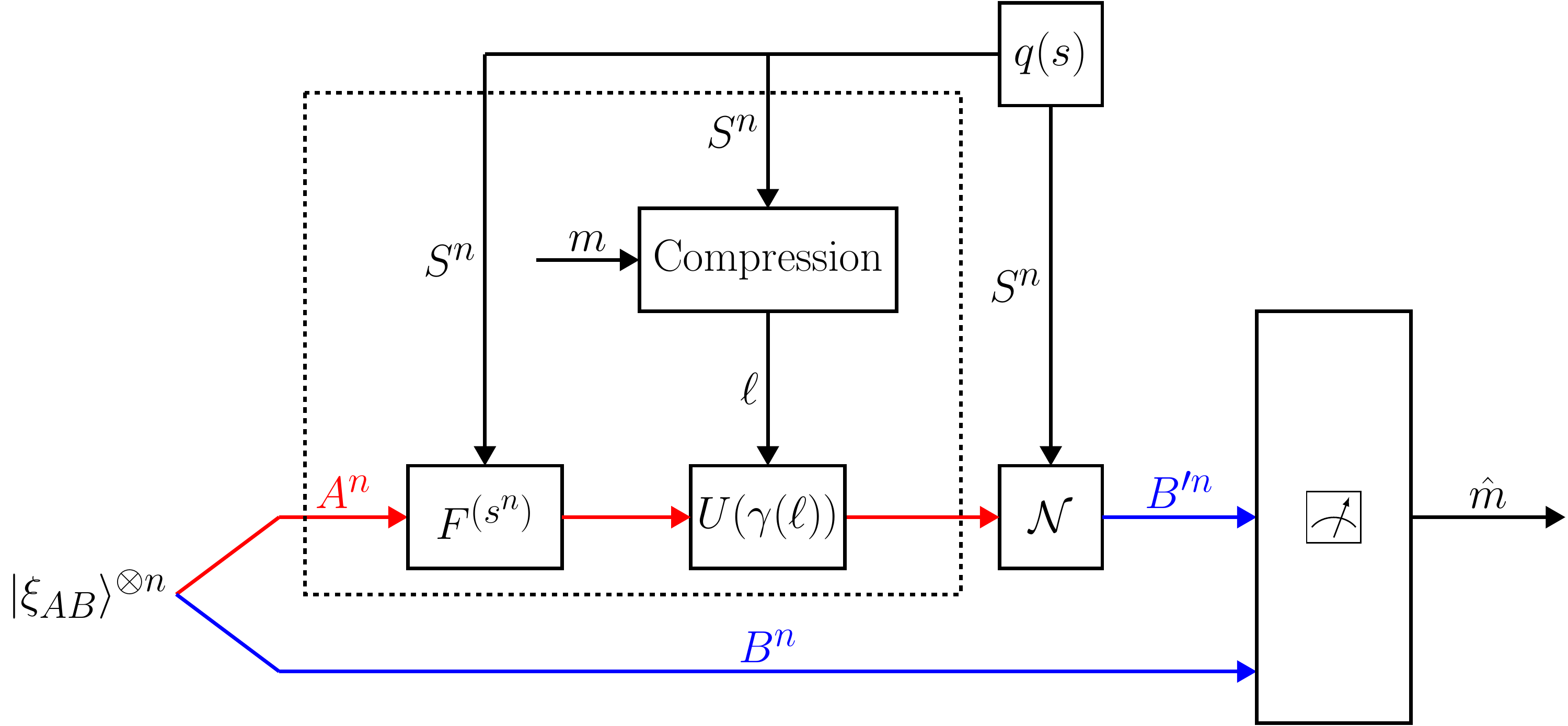}
\end{center}
\caption{
Coding scheme combining classical compression and generalized super-dense coding. The quantum systems of Alice and Bob are marked in red and blue, respectively. The blocks inside the dashed-line rectangle correspond to Alice's operations.
}
\label{fig:EAsiCodeDir}
\end{figure}
\subsubsection{Encoding and Decoding}
The coding scheme is depicted in Figure~\ref{fig:EAsiCodeDir}.
To send a message $m\in [1:2^{nR}]$, given a parameter sequence $s^n\in\Sset^n$, Alice performs the following.
\begin{enumerate}[(i)]
\item
 Find a sequence $x^n(\ell)\in\mathscr{B}(m)$ that is jointly typical with the parameter sequence, \ie $(s^n,x^n(\ell))\in\tset(p_{S,X})$. If there is none, choose an arbitrary $\ell$.

\item
 Apply the operators $F^{(s_1)},F^{(s_2)},\ldots,F^{(s_n)}$, and   $U(\gamma(\ell))$, which yields 
\begin{align}
|\varphi_{A^n B^n}^{\ell,s^n}\rangle\equiv (  U(\gamma(\ell) F^{(s^n)} \otimes\identity) |\xi_{AB} \rangle^{\otimes n} 
=(  U(\gamma(\ell)  \otimes\identity) |\varphi^{s^n}_{A^n B^n} \rangle
\end{align}
with $F^{(s^n)}\equiv F^{(s_1)}\otimes \cdots\otimes F^{(s_n)}$ and 
$|\varphi^{s^n}_{A^n B^n} \rangle\equiv |\varphi^{s_1}_{A B} \rangle\otimes\cdots\otimes
|\varphi^{s_n}_{A B} \rangle$ (see (\ref{eq:ABphis})).

\item Send the systems $A^n$ through the channel.
\end{enumerate}

Bob receives the systems $B'^n$ at state $\omega_{B'^n B^n}$ and applies a POVM $\{ \Lambda_\ell \}_{\ell\in [1:2^{n\tR}]}$, which will be specified later. Once Bob has a measurement result $\hat{\ell}$, he decodes the message as the corresponding sub-codebook.
That is, Bob declares the message to be $\hat{m}\in [1:2^{nR}]$ such that $x^n(\hat{\ell})\in\mathscr{B}(\hat{m})$.

\subsubsection{Code Properties}
First, we write the entangled states as a combination of maximally entangled states over the typical subspaces, and then we can use the following useful identities.
For a maximally entangled state $
| \Phi_{AB} \rangle = \frac{1}{\sqrt{D}} \sum_{j=0}^{D-1} |j\rangle_A \otimes |j\rangle_B 
$,
\begin{align}
\trace_B \left( \kb{ \Phi_{AB} }  \right) =\pi_A
\end{align}
where  $\pi_A=\frac{1}{D}\sum_{x\in\Xset} \kb{x}$ is the maximally mixed state.
Furthermore, for every state $\rho$ of the system $A$, 
\begin{align}
\frac{1}{D^2}\sum_{a=0}^{D-1} \sum_{b=0}^{D-1} \Sigma(a,b)\, \rho \,\Sigma^{\dagger}(a,b) 
=\pi_A 
\label{eq:piIden}
\end{align}
(see \eg \cite{BennettShorSmolin:99p}  \cite[Exercise 4.7.6]{Wilde:17b})).
 Another useful identity is the ``ricochet property" \cite[Eq. (17)]{HsiehDevetakWinter:08p},
\begin{align}
(U\otimes\identity)|\Phi_{AB}\rangle=(\identity\otimes U^T)|\Phi_{AB}\rangle \,.
\label{eq:ricochet}
\end{align}

Now, for every $s^n\in\Sset^n$, 
\begin{align}
|\varphi^{s^n}_{A^n, B^n} \rangle
=& \sum_{x^n\in\Xset^n} \sqrt{ p_{X^n|S^n}(x^n|s^n) }  |x^n\rangle \otimes |\psi_{x^n,s^n} \rangle
\end{align}
where $p_{X^n|S^n}(x^n|s^n)=\prod_{i=1}^n p_{X|S}(x_i|s_i)$ and $|\psi_{x^n} \rangle=|\psi_{x_1} \rangle\otimes|\psi_{x_2} \rangle\otimes\cdots\otimes|\psi_{x_n} \rangle$. As the space $\Xset^n$ can be partitioned into conditional type classes given $s^n$, we may write 
\begin{align}
|\varphi^{s^n}_{A^n, B^n} \rangle
=& \sum_{t\in\pSpace_n(\Xset)} \sum_{x^n\in \Tset_n(\hP|s^n)} \sqrt{ p_{X^n|S^n}(x^n|s^n) }  |x^n\rangle \otimes |\psi_{x^n,s^n} \rangle
\nonumber\\
=&  \sum_{t\in\pSpace_n(\Xset)} \sqrt{ p_{X^n|S^n}(x_t^n|s^n) } \sum_{x^n\in \Tset_n(t|s^n)}   |x^n\rangle \otimes |\psi_{x^n,s^n} \rangle
\end{align}
where $x_t^n$ is any sequence in the conditional type class $\Tset_n(t|s^n)$. Therefore, we have that 
\begin{align}
|\varphi^{s^n}_{A^n, B^n} \rangle
=&  \sum_{t\in\pSpace_n(\Xset)} \sqrt{ P(t|s^n)}   |\Phi_t \rangle \,,
\label{eq:phiKBsnT}
\end{align}
where
\begin{align}
& P(t|s^n)=d_t(s^n)\cdot p_{X^n|S^n}(x_t^n|s^n)  \,, \;\text{with $d_t(s^n)\equiv |\Tset_n(t|s^n)|$}
 \nonumber\\ 
& |\Phi_t \rangle=\frac{1}{\sqrt{d_t(s^n)}} \sum_{x^n\in \Tset_n(t|s^n)}   |x^n\rangle \otimes |\psi_{x^n,s^n} \rangle
\end{align}
We note that $P(t|s^n)$ is the conditional probability of the type $\Tset_n(t|s^n)$ for a classical random sequence
$X^n\sim p_{X^n|S^n=s^n}$.

Now, Alice applies the operator $U(\gamma(\ell))$ to the entangled states.
Since the state $|\Phi_t \rangle$ is  maximally entangled, we have by the ``ricochet property" 
(\ref{eq:ricochet}) that
\begin{align}
|\varphi_{A^n B^n}^{\gamma(\ell),s^n}\rangle\equiv (U(\gamma(\ell))\otimes \identity)
|\varphi^{s^n}_{A^n, B^n} \rangle =
(\identity\otimes U^T(\gamma(\ell)))|\varphi^{s^n}_{A^n, B^n} \rangle \,.
\label{eq:RicoCons}
\end{align}
By the same considerations, we also have that
\begin{align}
|\varphi^{s^n}_{A^n, B^n} \rangle = (F^{(s^n)}\otimes \identity) |\xi_{AB}\rangle^{\otimes n}=
(\identity \otimes (F^{(s^n)})^T ) |\xi_{AB}\rangle^{\otimes n} \,.
\end{align}
That is, Alice's unitary operations can be reflected and treated as if performed by Bob.

Bob then receives the systems $B'^n$ at state
\begin{align}
\rho^{\gamma(\ell)}_{ B'^n, B^n}=
&
\sum_{s^n\in\Sset^n} q^n(s^n)   (\channel^{(s^n)}\otimes\identity)  \left( 
\left( \kb{ \varphi_{A^n B^n}^{\gamma(\ell),s^n} }
\right)\right)
\label{eq:rhoG1}
\\
=&
\sum_{s^n\in\Sset^n} q^n(s^n)  (\channel^{(s^n)}\otimes\identity) \left( 
(\identity\otimes U^T(\gamma(\ell))) \kb{ \varphi_{A^n B^n}^{s^n} } (\identity\otimes U^*(\gamma(\ell)))
\right)
\label{eq:rhoG1c}
\end{align}
where the last line is due to  (\ref{eq:RicoCons}). 
Since a quantum channel is a linear map,  the above can be written as
\begin{align}
\rho^{\gamma}_{B'^n, B^n}
=& 
(\identity\otimes U^T(\gamma))  \left[ \sum_{s^n\in\Sset^n} q^n(s^n)
(\channel^{(s^n)}\otimes\identity)\left(  \kb{ \varphi_{A^n B^n}^{s^n} } \right) \right]
(\identity\otimes U^*(\gamma))
\nonumber\\
=& 
(\identity\otimes U^T(\gamma))  \omega_{B'B}^{\otimes n}
(\identity\otimes U^*(\gamma))
\label{eq:rhoG2}
\end{align}
where we have defined
\begin{align}
&\omega_{AB}^s= ( \identity\otimes (F^{(s)})^T )\kb{\xi_{AB}} ( \identity\otimes (F^{(s)})^* ) \\
&\omega_{SAB}=\sum_{s\in\Sset} q(s) \kb{ s } \otimes  \omega_{AB}^s  \label{eq:oSKB} \\
&\omega_{B'B}= ( \channel\otimes \identity)(\omega_{SAB}) \,.
\label{eq:oBpB}
\end{align}

\subsubsection{Packing Lemma Requirements}
Next, we use the quantum packing lemma. Consider the ensemble 
$\left\{ p(\gamma)=\frac{1}{|\Gamma|},  \rho^{\gamma}_{B'^n, B^n} \right\}$, for which the expected density operator is 
\begin{align}
\sigma_{B'^n,B^n}=\frac{1}{|\Gamma|}\sum_{\gamma\in\Gamma} \rho^{\gamma}_{B'^n, B^n} \,.
\label{eq:SigmaBpB}
\end{align}
Define the code projector and the codeword projectors by
\begin{align}
 \Pi \equiv& \Pi^{\delta}(\omega_{B'})\otimes \Pi^{\delta}(\omega_{B})
\label{eq:CodeProjDir}
\\
 \Pi_\gamma \equiv& (\identity\otimes U^T(\gamma)) \Pi^{\delta}(\omega_{B'B}) (\identity\otimes U^*(\gamma)) \,,\; \text{for 
$\gamma\in\Gamma$}
\end{align}
where $\Pi^{\delta}(\omega_{B'B})$, $\Pi^{\delta}(\omega_{B'})$ and $\Pi^{\delta}(\omega_{B})$ are the projectors onto the $\delta$-typical subspaces associated with the states $\omega_{B'B}$, $\omega_{B'}=\trace_B(\omega_{B'B})$ and
$\omega_{B}=\trace_{B'}(\omega_{B'B})$, respectively (see (\ref{eq:oBpB})).
Now, we verify that the assumptions of Lemma~\ref{lemm:Qpacking} hold with respect to the ensemble and the projectors above.

First, we show that $\trace(\Pi \rho^{\gamma}_{B'^n, B^n})\geq\, 1-\alpha$, where $\alpha>0$ is arbitrarilly small. Defining 
$\check{P}=\identity-P$, we have that
\begin{align}
\Pi= (\identity-\check{\Pi}^{\delta}(\rho_{B'}))\otimes (\identity- \check{\Pi}^{\delta}(\rho_{B}) )
\succeq (\identity\otimes\identity)-(\check{\Pi}^{\delta}(\rho_{B'})\times\identity)-(\identity\otimes\check{\Pi}^{\delta}(\rho_{B}))
\end{align}
hence,
\begin{align}
\trace(\Pi \rho^{\gamma}_{B'^n, B^n})\geq& 1
-\trace\left( (\check{\Pi}^{\delta}(\rho_{B'})\otimes\identity) \rho^{\gamma}_{B'^n, B^n} \right)
-\trace\left( (\identity\otimes\check{\Pi}^{\delta}(\rho_{B})) \rho^{\gamma}_{B'^n, B^n} \right)
\nonumber\\
=& 1
-\trace\left( \check{\Pi}^{\delta}(\rho_{B'}) \rho^{\gamma}_{B'^n} \right)
-\trace\left( \check{\Pi}^{\delta}(\rho_{B}) \rho^{\gamma}_{ B^n} \right) \,.
\label{eq:ReqPackDir1}
\end{align}
The first trace term in the RHS of (\ref{eq:ReqPackDir1}) equals $\trace\left( \check{\Pi}^{\delta}(\rho_{B'}) \omega_{B'}^{\otimes n} \right)$ by 
(\ref{eq:rhoG2}), and the last term equals $\trace\left( \check{\Pi}^{\delta}(\rho_{B})  \omega_{B}^{\otimes n}  \right)$ by (\ref{eq:rhoG1})
and (\ref{eq:oSKB}). Therefore, we have by (\ref{eq:UnitT}) that 
\begin{align}
\trace(\Pi \rho^{\gamma}_{B'^n, B^n})\geq& 
1
-\trace\left( \check{\Pi}^{\delta}(\rho_{B'}) \omega_{B'}^{\otimes n} \right)
-\trace\left( \check{\Pi}^{\delta}(\rho_{B}))  \omega_{B}^{\otimes n}  \right)
\nonumber\\
\geq& 1-2\eps \,.
\end{align}
Similarly, the second requirement of the packing lemma  holds since
\begin{align}
\trace(\Pi_\gamma \rho^{\gamma}_{B'^n, B^n})=&\trace\left[ (\identity\otimes U^T(\gamma)) \Pi^{\delta}(\omega_{B'B}) (\identity\otimes U^*(\gamma))
(\identity\otimes U^T(\gamma))  \omega_{B'B}^{\otimes n}
(\identity\otimes U^*(\gamma))
  \right]
	\nonumber\\
	=&\trace( \Pi^{\delta}(\omega_{B'B})\omega_{B'B}^{\otimes n}) \geq 1-\eps 
\end{align}
where the second equality follows from the cyclicity of the trace and the fact that $U^* U^T=(U U^\dagger)^*=\identity$ for a unitary operator, and the last inequality is due to (\ref{eq:UnitT}).

Moving to the third requirement in Lemma~\ref{lemm:Qpacking},
\begin{align}
\trace(\Pi_\gamma)=\trace\left( (\identity\otimes U^T(\gamma)) \Pi^{\delta}(\omega_{B'B}) (\identity\otimes U^*(\gamma)) \right)
=\trace( \Pi^{\delta}(\omega_{B'B}) ) \leq 2^{n(H(\omega_{B'B})+c\delta)}
\end{align}
where the second equality holds by cyclicity of the trace and the last inequality is due to (\ref{eq:Pidim}).
It is left to verify that the last requirement of the packing lemma holds, \ie
$\Pi \sigma_{B'^n,B^n} \Pi \preceq\, 2^{-n(H(\sigma_{B'})+H(\sigma_B)-\alpha)} \Pi$. To this end, observe that by (\ref{eq:rhoG1c}) and 
(\ref{eq:SigmaBpB}),
\begin{align}
\sigma_{B'^n,B^n}= \sum_{s^n\in\Sset^n} q^n(s^n) (\channel^{(s^n)}\otimes\identity) \tau^{s^n}_{A^n,B^n}
\label{eq:sigmaxiBp}
\end{align}
where we have defined
\begin{align}
\tau^{s^n}_{A^n,B^n}\equiv \frac{1}{|\Gamma|} \sum_{\gamma\in\Gamma} (\identity\otimes U^T(\gamma))  \kb{ \varphi_{A^n B^n}^{s^n} } 
(\identity\otimes U^*(\gamma)) \,.
\label{eq:xisn}
\end{align}
Then, by (\ref{eq:RicoCons}) along with (\ref{eq:Vt})-(\ref{eq:Ugamma}),
\begin{align}
\tau^{s^n}_{A^n,B^n}
=&  \frac{1}{|\Gamma|} \sum_{\gamma\in\Gamma} (\identity\otimes U^T(\gamma)) \left(   \sum_{t} \sqrt{ P(t|s^n)}
   |\Phi_t \rangle  \right)
\left(   \sum_{t'} \sqrt{ P(t'|s^n)}   \langle\Phi_{t'} | \right)  
(\identity\otimes U^*(\gamma)) 
\nonumber\\
=&  \frac{1}{|\Gamma|} \sum_{\gamma\in\Gamma} 
\left(   \sum_{t} \sqrt{ P(t|s^n)} (-1)^{c_t(\gamma)}  (\identity\otimes \Sigma^T_{a_t(\gamma),b_t(\gamma)})
   |\Phi_t \rangle  \right)
	\nonumber\\
	&\quad
\left(   \sum_{t'} \sqrt{ P(t'|s^n)}  (-1)^{c_{t'}(\gamma)}   \langle\Phi_{t'} |  (\identity\otimes \Sigma^*_{a_{t'}(\gamma),b_{t'}(\gamma)}) \right)
\,.
\label{eq:xisn1}
\end{align}
For $t'=t$, the expression above becomes
\begin{align}
 &\frac{1}{|\Gamma|} \sum_{\gamma\in\Gamma}   \sum_{t}  P(t|s^n)
 (\identity\otimes \Sigma^T_{a_t(\gamma),b_t(\gamma)})
  \kb{ \Phi_t }   (\identity\otimes  \Sigma^*_{a_{t}(\gamma),b_{t}(\gamma)}) 
	\nonumber\\
	=& \sum_{t}  P(t|s^n) \left[ \frac{1}{D_t^2} \sum_{a_t,b_t}  (\identity\otimes \Sigma^T_{a_t,b_t})
  \kb{ \Phi_t }   (\identity\otimes  \Sigma^*_{a_{t},b_{t}}) \right]
	= \sum_{t}  P(t|s^n) \pi^t_{A^n} \times \pi^t_{B^n}
	\label{eq:xisn2}
\end{align}
with
\begin{align}
\pi^t_{A^n}\equiv\frac{\Pi_{A^n}(t)}{\trace(\Pi_{A^n}(t))} \,,\;
\pi^t_{B^n}\equiv \frac{\Pi_{B^n}(t)}{\trace(\Pi_{B^n}(t))}
\label{eq:pitKBn}
\end{align}
where $\Pi_{A^n}(t)$ is the projector of type $t$ as defined in (\ref{eq:ProjType}).
 The last equality in (\ref{eq:xisn2}) follows from (\ref{eq:piIden}). On the other hand, for $t'\neq t$, 
\begin{align}
  &\frac{1}{|\Gamma|} \sum_{\gamma\in\Gamma} 
   \sum_{t} \sum_{t'\neq t} \sqrt{ P(t|s^n)  P(t'|s^n)} 
	\frac{1}{4D_t^2 D_{t'}^2} \sum_{c_t,c_{t'}\in \{0,1\}}
	(-1)^{c_t+c_{t'}} \sum_{a_t,a_{t'},b_t,b_{t'}}  (\identity\otimes \Sigma^T_{a_t,b_t})
   |\Phi_t \rangle  
    \langle\Phi_{t'} |  (\identity\otimes \Sigma^*_{a_{t'},b_{t'}}) 
		= 0
\,.
\label{eq:xisn3}
\end{align}
We deduce from (\ref{eq:xisn1})-(\ref{eq:xisn3}) that $
\tau^{s^n}_{A^n,B^n}= \sum_{t}  P(t|s^n) \pi^t_{A^n} \otimes \pi^t_{B^n}
$. Plugging this into (\ref{eq:sigmaxiBp}) yields
\begin{align}
\sigma_{B'^n,B^n}= \sum_{s^n\in\Sset^n} q^n(s^n)  \sum_{t}  P(t|s^n) \channel^{(s^n)}( \pi^t_{A^n})\otimes  \pi^t_{B^n} \,.
\end{align}

Now, we use the formula above in order to show that the last requirement in Lemma~\ref{lemm:Qpacking} holds. Consider that
\begin{align}
\Pi \sigma_{B'^n,B^n} \Pi =& (\Pi^{\delta}(\omega_{B'})\otimes \Pi^{\delta}(\omega_{B})) \sigma_{B'^n,B^n}
 (\Pi^{\delta}(\omega_{B'})\otimes \Pi^{\delta}(\omega_{B}))
\nonumber\\
=& \sum_{s^n\in\Sset^n} q^n(s^n)  \sum_{t}  P(t|s^n) 
\big( \Pi^{\delta}(\omega_{B'})\channel^{(s^n)}( \pi^t_{A^n}) \Pi^{\delta}(\omega_{B'}) \big)\otimes
\big( \Pi^{\delta}(\omega_{B}) \pi^t_{B^n} \Pi^{\delta}(\omega_{B}) \big) \,.
\end{align}
Using (\ref{eq:pitKBn}), this can be bounded by
\begin{align}
\Pi \sigma_{B'^n,B^n} \Pi 
=& \sum_{s^n\in\Sset^n} q^n(s^n)  \sum_{t}  P(t|s^n) 
\big( \Pi^{\delta}(\omega_{B'})\channel^{(s^n)}( \pi^t_{A^n}) \Pi^{\delta}(\omega_{B'}) \big)\otimes
\big( \Pi^{\delta}(\omega_{B}) \frac{\Pi_{B^n}(t)}{\trace(\Pi_{B^n}(t))} \Pi^{\delta}(\omega_{B}) \big)
\nonumber\\
\preceq&
2^{-n(H(\omega_B)+\eps_1)} \sum_{s^n\in\Sset^n} q^n(s^n)  \sum_{t}  P(t|s^n) 
\big( \Pi^{\delta}(\omega_{B'})\channel^{(s^n)}( \pi^t_{A^n}) \Pi^{\delta}(\omega_{B'}) \big)\otimes
 \Pi^{\delta}(\omega_{B}) 
\end{align}
with arbitrarily small $\eps_1>0$, following (\ref{eq:Tsize}) and the fact that
$\Pi^{\delta}(\omega_{B})\Pi_{B^n}(t)\Pi^{\delta}(\omega_{B}) \preceq \Pi^{\delta}(\omega_{B})$.
By linearity, this can also be written as
\begin{align}
\Pi \sigma_{B'^n,B^n} \Pi 
\preceq&
2^{-n(H(\omega_B)+\eps_1)} 
 \Pi^{\delta}(\omega_{B'}) \left[ \sum_{s^n\in\Sset^n} q^n(s^n)   \channel^{(s^n)}\left( \sum_{t}  P(t|s^n) \pi^t_{A^n} \right) \right]
 \Pi^{\delta}(\omega_{B'})  \otimes
 \Pi^{\delta}(\omega_{B}) 
\nonumber\\
=&
2^{-n(H(\omega_B)+\eps_1)} 
 \Pi^{\delta}(\omega_{B'}) \left[ \sum_{s^n\in\Sset^n} q^n(s^n)    \channel^{(s^n)}\left( \omega^{s^n}_{A^n} \right) \right]
 \Pi^{\delta}(\omega_{B'})  \otimes
 \Pi^{\delta}(\omega_{B}) 
\end{align}
(see (\ref{eq:phiKBsnT})). Since the expression in the square brackets equals $\omega_{B'}^{\otimes n}$ (see (\ref{eq:oBpB})), we have by 
 (\ref{eq:rhonProjIneq})  that 
\begin{align}
\Pi \sigma_{B'^n,B^n} \Pi \preceq 2^{-n(H(\omega_B')+H(\omega_B)+\eps_1+\eps_2)}  \Pi^{\delta}(\omega_{B'})\otimes  \Pi^{\delta}(\omega_{B})
=2^{-n(H(\omega_B')+H(\omega_B)+\eps_1+\eps_2)} \Pi 
\end{align}
with arbitrarily small $\eps_2>0$,
where the last equality follows from the definition of $\Pi$ in (\ref{eq:CodeProjDir}). It follows that all of the requirements of the packing lemma are satisfied. 

Hence, by Lemma~\ref{lemm:Qpacking}, there exist deterministic vectors $\gamma(\ell)$, $\ell\in [1:2^{n\tR}]$, and  a POVM 
$\{ \Lambda_{\ell} \}_{\ell\in [1:2^{n\tR}]}$ such that 
\begin{align}
  \trace\left( \Lambda_{\ell} \rho^{\gamma(\ell)}_{ B'^n, B^n} \right)  \geq 1-2^{-n[ I(B';B)_\omega-\tR-\eps']}
	\label{eq:packDirIneq}
\end{align}
for all 
$\ell\in [1:2^{n\tR}]$, where $\eps'$ is arbitrarily small.

\subsubsection{Error Probability Analysis}
 Observe that Bob can only decode the message $m$ correctly if Alice chooses $\ell$ such that 
$\ell\in\mathscr{B}(m)$. Due to the symmetry, we may assume without loss of generality that Alice chose the message $m=1$ and compressed the state sequence using 
$\ell=1$. Hence, the error event is bounded by the union of the following events
\begin{align}
\mathscr{F}_1 =&\{  (S^n,X^n(\ell'))\notin \tset(p_{S,X}) \;\text{for all $\ell'\in [1:2^{n(\tR-R)}]$} \} \\
\mathscr{F}_2=& \{  \hat{\ell}\neq 1 
\} \,.
\end{align}
Thus, by the union of events bound
\begin{align}
P_{e|m=1}^{(n)}(\Eset,\phi_{KB},\Lambda)\leq& \prob{\mathscr{F}_1\cup\mathscr{F}_2}\leq
\prob{ \mathscr{F}_1  }+
\prob{  \mathscr{F}_2  }
\nonumber\\
=& \prob{ \mathscr{F}_1  }+
\trace\left(   (\identity-\Lambda_\ell) \rho^{\gamma(\ell)}_{ B'^n, B^n} \right) \,,
\label{eq:Edir1}
\end{align}
where the conditioning on $m=1$ and $\ell=1$ is omitted for convenience of notation. 
By the classical covering lemma (see Lemma~\ref{lemm:covering}), we have that
$
\prob{ \mathscr{F}_1  }\leq \exp\big( -2^{n(\tR-R- I(X;S)-\eps' )} \big)
$. We also have that $ I(X;S)= I(B;S)_{\varphi}=I(B;S)_{\omega}$ by 
(\ref{eq:IBS}) and (\ref{eq:oSKB}).
Hence, the first term in the RHS of (\ref{eq:Edir1}) tends to zero as $n\rightarrow\infty$ provided that
\begin{align}
R<
\tR-I(B;S)_{\omega}-\eps' \,.
\end{align}
Based on (\ref{eq:packDirIneq}), the second term in the RHS of (\ref{eq:Edir1}) is bounded by $2^{-n[ I(B';B)_\omega-\tR-\eps_n(\alpha)]}$, which tends to zero if
\begin{align}
\tR< I(B';B)_\omega-\eps' \,,
\end{align}
for sufficiently large $n$ and small $\alpha>0$.
Therefore, the probability of error tends to zero as $n\rightarrow\infty$ for $\tR=R+\eps'$ and $R<I(B;B')_\omega-I(B;S)_{\omega}-3\eps'$.

Now, consider the systems $S,A_1,A_1',B_1$ at state
\begin{align}
&|\varphi^s_{  A_1 A_1' }\rangle= (F_1^{(s)}\otimes\identity  )  | \xi_{A_1,A_1'}  \rangle  \label{eq:oSKB0} \\
&\omega_{A_1 S  A_1' }=\sum_{s\in\Sset} q(s) \kb{ s } \otimes  \varphi^s_{  A_1 A_1' } \\
&\omega_{ A_1 B_1}= (\identity\otimes \channel)(\omega_{ A_1 S A_1'}) \,.
\label{eq:oKAB10}
\end{align}
Observe that those are the same relations as in (\ref{eq:oBpB}) where  $A_1'$,  $A_1$ and $B_1$
take place with  $A$,   $B$, and $B'$, respectively, with $F_1^{(s)}\equiv (F^{(s)})^T$ for $s\in\Sset$.
Thus, the probability of error tends to zero as $n\rightarrow\infty$ provided that  $R<I(A_1;B_1)_\omega-I(A_1;S)_{\omega}-3\eps'$.
This completes the proof of the direct part.

\begin{remark}
At a first glance, it may seem that we can modify the proof above to prove  Theorem~\ref{theo:mainC} for causal CSI
by simply removing the compression stage of the encoding procedure, and continuing the analysis without conditioning on the state sequence. However, such coding scheme would still violate the causality requirement, since Alice cannot apply the operator $U(\gamma)$ to the entire sequence of input systems (see Figure~\ref{fig:EAsiCodeDir}). Instead, in the proof of Theorem~\ref{theo:mainC} in 
Appendix~\ref{app:mainC}, Alice applies the encoding operations in a \emph{reversed order}, \ie first $U(\gamma)$ is applied to 
a sequence of auxiliary systems $K^n$, which do not depend on the state sequence, and only then $\Fset^{(s_i)}$ are applied
(see Figure~\ref{fig:EAsiCodeDirC}).
\label{rem:CreversedEop}
\end{remark}

\subsection{Converse Proof}
Consider the converse part. Suppose that Alice and Bob are trying to distribute randomness. An upper bound on the rate at which Alice can distribute randomness to Bob also serves as an upper bound on the rate at which they can communicate. In this task, Alice and Bob share an entangled state $\Psi_{T_A^n T_B^n}$. Alice first prepares the maximally corrleated state
\begin{align}
\overline{\Phi}_{MM'} \equiv \frac{1}{2^{nR}}\sum_{m=1}^{2^{nR}} \kb{ m } \otimes \kb{ \phi_m } \,.
\end{align}
locally.
 Then, Alice applies an encoding channel $\Eset^{s^n}_{M'T_A^n \rightarrow A'^n}$ to the classical system $M'$ and her share $T_A^n$ of the entangled state $\Psi_{T_A^n T_B^n}$. 
The resulting state is $\omega_{S^n M A'^n T_B^n}=\sum_{s^n\in\Sset^n} q^n(s^n) \kb{ s^n }\otimes 
\rho^{s^n}_{ M A'^n T_B^n}$, with
\begin{align}
\rho^{s^n}_{ M  A'^n T_B^n }\equiv (\identity\otimes\Eset^{s^n}\otimes\identity)( \overline{\Phi}_{MM'}\otimes \Psi_{T_A^n T_B^n} ) \,.
\end{align}
After Alice sends the systems $A'^n$ through the channel, Bob receives the systems $B^n$ at state
$\omega_{S^n M A'^n T_B^n}=\sum_{s^n\in\Sset^n} q^n(s^n) \kb{ s^n }\otimes 
\rho^{s^n}_{ M B^n T_B^n}$, with
\begin{align}
\rho^{s^n}_{M B^n T_B^n}\equiv (\identity\otimes\channel^{(s^n)}\otimes\identity) (\rho^{s^n}_{ M  A'^n T_B^n }) \,.
\end{align}
Then, Bob performs a decoding channel $\Dset_{B^n T_B^n\rightarrow \hM}$, producing $\omega_{S^n M \hM}'=\sum_{s^n\in\Sset^n} q^n(s^n) \kb{ s^n }\otimes 
\rho^{s^n}_{ M \hM}$ with
\begin{align}
\rho^{s^n}_{ M \hM}\equiv (\identity\otimes\Dset)(\rho^{s^n}_{M B^n T_B^n})
\label{eq:DecConv1}
\end{align}

Consider a sequence of codes $(\Eset_n^{s^n},\Psi_n,\Dset_n)$ for randomness distribution, such that
\begin{align}
\frac{1}{2} \norm{ \omega_{M\hM} -\overline{\Phi}_{MM'} }_1 \leq \alpha_n \,,
\end{align}
where $\omega_{M\hM}$ is the reduced density operator of $\omega_{S^n M \hM}$ and while $\alpha_n$ tends to zero as $n\rightarrow\infty$.
By the Alicki-Fannes-Winter inequality \cite{AlickiFannes:04p,Winter:16p} \cite[Theorem 11.10.3]{Wilde:17b}, this implies that
\begin{align}
|H(M|\hM)_\omega - H(M|M')_{\overline{\Phi}} |\leq n\eps_n
\label{eq:AFW}
\end{align}
while $\eps_n$ tends to zero as $n\rightarrow\infty$.
Now, observe that $H(\overline{\Phi}_{M M'})=H(\overline{\Phi}_{M})=H(\overline{\Phi}_{ M'})=nR$, hence $I(M;\hM)_{\overline{\Phi}}=nR$.
Also,   $H(\omega_{M})=H(\overline{\Phi}_{ M})=nR$ implies that 
$I(M;M')_{\overline{\Phi}} - I(M;\hM)_{\omega}= H(M|\hM)_\omega - H(M|M')_{\overline{\Phi}}$. Therefore, by (\ref{eq:AFW}),
\begin{align}
nR=&I(M;\hM)_{\overline{\Phi}} \nonumber\\
\leq& I(M;\hM)_{\omega}+n\eps_n \nonumber\\
\leq& I(M;T_B^n,B^n)_{\omega}+n\eps_n 
\label{eq:ConvIneq1}
\end{align}
where the last line follows from (\ref{eq:DecConv1}) and the quantum data processing inequality \cite[Theorem 11.5]{NielsenChuang:02b}.

As in the classical case, the chain rule for the quantum mutual information states that 
$I(A;B,C)_\sigma=I(A;B)_\sigma+I(A;C|B)_\sigma$ for all $\sigma_{ABC}$ (see \eg \cite[Property 11.7.1]{Wilde:17b}). 
As a straightforward consequence, this leads to the Cisz\'ar sum identity,
\begin{align}
\sum_{i=1}^n I(A_{i+1}^n;B_i|B^{i-1})_\sigma=\sum_{i=1}^n I(B^{i-1};A_i|A_{i+1}^n)_\sigma
\label{eq:CsiszarIdentity}
\end{align}
for every sequence of systems $A^n$ and $B^n$.
Returning to (\ref{eq:ConvIneq1}), we apply the chain rule and rewrite the inequality as
\begin{align}
nR \leq& I(T_B^n,M;B^n)_{\omega}+I(M;T_B^n)_{\omega}-I(T_B^n;B^n)_{\omega}+n\eps_n \nonumber\\
\leq& I(T_B^n,M;B^n)_{\omega}+I(M;T_B^n)_{\omega}+n\eps_n \nonumber\\
=& I(T_B^n,M;B^n)_{\omega}+n\eps_n 
\label{eq:ConvIneq2}
\end{align}
where the equality holds since the systems $M$ and $T_B^n$ are in a product state. The chain rule further 
implies that 
\begin{align}
I(T_B,M;B^n)_{\omega}=&\sum_{i=1}^n I(T_{B}^n,M;B_i| B^{i-1})_\omega 
\nonumber\\
\leq& \sum_{i=1}^n I(T_B^n,M,B^{i-1};B_i)_\omega \nonumber\\
=& \sum_{i=1}^n I(T_B^n,M,B^{i-1},S_{i+1}^n;B_i)_\omega -\sum_{i=1}^n I(B_i;S_{i+1}^n|T_B^n,M,B^{i-1})_\omega \nonumber\\
=& \sum_{i=1}^n I(T_B^n,M,B^{i-1},S_{i+1}^n;B_i)_\omega -\sum_{i=1}^n I(B^{i-1};S_i|T_B^n,M,S_{i+1}^n)_\omega
\label{eq:ConvIneq3}
\end{align}
where the last line follows from the quantum version of the Csisz\'ar sum identity in (\ref{eq:CsiszarIdentity}).
Since the systems $S_i$ and $(T_B^n,M,S_{i+1}^n)$ are in a product state, $I(B^{i-1};S_i|T_B^n,M,S_{i+1}^n)_\omega=
I(T_B^n,M,S_{i+1}^n,B^{i-1};S_i)_\omega$. 
Defining $K_i=(M,M',S^{i-1},S_{i+1}^n,T_A^n,T_B^n)$ and a quantum channel $\Fset^{(s_i)}_{K_i\rightarrow A_i}$ such that 
$A_i=(M,B^{i-1},S_{i+1}^n,T_B^n)$, we have by (\ref{eq:ConvIneq2}) and (\ref{eq:ConvIneq3}) that
\begin{align}
R-\eps_n\leq \frac{1}{n}\sum_{i=1}^n [ I(A_i;B_i)_\omega-I(A_i;S_i)_\omega ] 
\leq \max_{\theta_{KA'}\,,\; \Fset^{(s)}_{K\rightarrow A}} [ I(A;B)_\omega-I(A;S)_\omega ] 
\,.
\end{align}
This concludes the proof of Theorem~\ref{theo:mainNC}.
\qed 

\end{appendices}

 
\bibliography{references2}{}

\ifdefined\bibstar\else\newcommand{\bibstar}[1]{}\fi
\begin{thebibliography}{73}
\providecommand{\natexlab}[1]{#1}
\providecommand{\url}[1]{\texttt{#1}}
\expandafter\ifx\csname urlstyle\endcsname\relax
  \providecommand{\doi}[1]{doi: #1}\else
  \providecommand{\doi}{doi: \begingroup \urlstyle{rm}\Url}\fi

\bibitem[Alicki and Fannes(2004)]{AlickiFannes:04p}
R.~Alicki and M.~Fannes.
\newblock Continuity of quantum conditional information.
\newblock \emph{J. Phys. A: Math. General}, 37\penalty0 (5):\penalty0 L55--L57,
  Jan 2004.

\bibitem[Anshu et~al.(2017)Anshu, Jain, and Warsi]{AnshuJainWarsi:17a}
A.~Anshu, R.~Jain, and N.~A. Warsi.
\newblock One shot entanglement assisted classical and quantum communication
  over noisy quantum channels: A hypothesis testing and convex split approach.
\newblock \emph{\textup{\texttt{arXiv:1702.01940}}}, 2017.

\bibitem[Anshu et~al.(2019)Anshu, Jain, and Warsi]{AnshuJainWarsi:19p}
A.~Anshu, R.~Jain, and N.~A. Warsi.
\newblock On the near-optimality of one-shot classical communication over
  quantum channels.
\newblock \emph{J. Math. Phys.}, 60\penalty0 (1):\penalty0 012204, 2019.

\bibitem[Barnum et~al.(1998)Barnum, Nielsen, and
  Schumacher]{BarnumNielsenSchumacher:98p}
H.~Barnum, M.~A. Nielsen, and B.~Schumacher.
\newblock Information transmission through a noisy quantum channel.
\newblock \emph{Phys. Rev. A}, 57\penalty0 (6):\penalty0 4153, June 1998.

\bibitem[Becerra et~al.(2015)Becerra, Fan, and Migdall]{BecerraFanMigdall:15p}
F.~E. Becerra, J.~Fan, and A.~Migdall.
\newblock Photon number resolution enables quantum receiver for realistic
  coherent optical communications.
\newblock \emph{Nature Photonics}, 9\penalty0 (1):\penalty0 48, 2015.

\bibitem[Bennett and Brassard(2014)]{BennettBrassard:14p}
C.~H. Bennett and G.~Brassard.
\newblock Quantum cryptography: public key distribution and coin tossing.
\newblock \emph{Theor. Comput. Sci.}, 560\penalty0 (12):\penalty0 7--11, 2014.

\bibitem[Bennett et~al.(1999)Bennett, Shor, Smolin, and
  Thapliyal]{BennettShorSmolin:99p}
C.~H. Bennett, P.~W. Shor, J.~A. Smolin, and A.~V. Thapliyal.
\newblock Entanglement-assisted classical capacity of noisy quantum channels.
\newblock \emph{Phys. Rev. Lett.}, 83\penalty0 (15):\penalty0 3081, Oct 1999.

\bibitem[{Bennett} et~al.(2002){Bennett}, {Shor}, {Smolin}, and
  {Thapliyal}]{BennettShorSmolin:02p}
C.~H. {Bennett}, P.~W. {Shor}, J.~A. {Smolin}, and A.~V. {Thapliyal}.
\newblock Entanglement-assisted capacity of a quantum channel and the reverse
  shannon theorem.
\newblock \emph{IEEE Trans. Inf. Theory}, 48\penalty0 (10):\penalty0
  2637--2655, Oct 2002.

\bibitem[Boche et~al.(2016)Boche, Cai, and N{\"o}tzel]{BocheCaiNotzel:16p}
H.~Boche, N.~Cai, and J.~N{\"o}tzel.
\newblock The classical-quantum channel with random state parameters known to
  the sender.
\newblock \emph{J. Physics A: Math. and Theor.}, 49\penalty0 (19):\penalty0
  195302, April 2016.

\bibitem[Bouwmeester and Zeilinger(2000)]{BouwmeesterZeilinger:00b}
D.~Bouwmeester and A.~Zeilinger.
\newblock The physics of quantum information: basic concepts.
\newblock In \emph{The physics of quantum information}, pages 1--14. Springer,
  2000.

\bibitem[Cacciapuoti et~al.(2019)Cacciapuoti, Caleffi, Van~Meter, and
  Hanzo]{CCVH:19a}
A.~S. Cacciapuoti, M.~Caleffi, R.~Van~Meter, and L.~Hanzo.
\newblock When entanglement meets classical communications: Quantum
  teleportation for the quantum internet.
\newblock \emph{\textup{\texttt{arXiv:1907.06197}}}, 2019.

\bibitem[Chen and Wornell(2001)]{ChenWornell:01p}
B.~Chen and G.~W. Wornell.
\newblock Quantization index modulation: A class of provably good methods for
  digital watermarking and information embedding.
\newblock \emph{IEEE Trans. Inf. Theory}, 47\penalty0 (4):\penalty0 1423--1443,
  May 2001.

\bibitem[Cheng et~al.(2018)Cheng, Hanson, Datta, and Hsieh]{CHDH:18a}
H.~C. Cheng, E.~P. Hanson, N.~Datta, and M.~H. Hsieh.
\newblock Duality between source coding with quantum side information and cq
  channel coding.
\newblock \emph{\textup{\texttt{arXiv:1809.11143}}}, 2018.

\bibitem[Cheng et~al.(2019)Cheng, Hanson, Datta, and Hsieh]{CHDH:19c}
H.~C. Cheng, E.~P. Hanson, N.~Datta, and M.~H. Hsieh.
\newblock Duality between source coding with quantum side information and cq
  channel coding.
\newblock In \emph{Proc. IEEE Int. Symp. Inf. Theory (ISIT'2019)}, pages
  1142--1146, Paris, France, July 2019.

\bibitem[Choudhuri et~al.(2013)Choudhuri, Kim, and
  Mitra]{ChoudhuriKimMitra:13p}
C.~Choudhuri, Y.~H. Kim, and U.~Mitra.
\newblock Causal state communication.
\newblock \emph{IEEE Trans. Inf. Theory}, 59\penalty0 (6):\penalty0 3709--3719,
  June 2013.

\bibitem[Csisz{\'a}r and K{\"o}rner(2011)]{CsiszarKorner:82b}
I.~Csisz{\'a}r and J.~K{\"o}rner.
\newblock \emph{Information Theory: Coding Theorems for Discrete Memoryless
  Systems}.
\newblock Cambridge University Press, 2 edition, 2011.

\bibitem[{Datta} and {Hsieh}(2013)]{DattaHsieh:13p}
N.~{Datta} and M.~{Hsieh}.
\newblock One-shot entanglement-assisted quantum and classical communication.
\newblock \emph{IEEE Trans. Inf. Theory}, 59\penalty0 (3):\penalty0 1929--1939,
  March 2013.

\bibitem[Datta et~al.(2018)Datta, Hirche, and Winter]{DattaHircheWinter:18a}
N.~Datta, C.~Hirche, and A.~Winter.
\newblock Convexity and operational interpretation of the quantum information
  bottleneck function.
\newblock \emph{\textup{\texttt{arXiv:1810.03644}}}, 2018.

\bibitem[Datta et~al.(2019)Datta, Hirche, and Winter]{DattaHircheWinter:19c}
N.~Datta, C.~Hirche, and A.~Winter.
\newblock Convexity and operational interpretation of the quantum information
  bottleneck function.
\newblock In \emph{Proc. IEEE Int. Symp. Inf. Theory (ISIT'2019)}, pages
  1157--1161, Paris, France, July 2019.

\bibitem[Devetak(2005)]{Devetak:05p}
I.~Devetak.
\newblock The private classical capacity and quantum capacity of a quantum
  channel.
\newblock \emph{IEEE Trans. Inf. Theory}, 51\penalty0 (1):\penalty0 44--55,
  2005.

\bibitem[Devetak and Shor(2005)]{DevetakShor:05p}
I.~Devetak and P.~W. Shor.
\newblock The capacity of a quantum channel for simultaneous transmission of
  classical and quantum information.
\newblock \emph{Commun. in Math. Phys.}, 256\penalty0 (2):\penalty0 287--303,
  June 2005.

\bibitem[Devetak and Winter(2003)]{DevetakWinter:03p}
I.~Devetak and A.~Winter.
\newblock Classical data compression with quantum side information.
\newblock \emph{Phys. Rev. A}, 68:\penalty0 042301, Oct 2003.

\bibitem[{Devetak} et~al.(2008){Devetak}, {Harrow}, and
  {Winter}]{DevetakHarrowWinter:08p}
I.~{Devetak}, A.~W. {Harrow}, and A.~J. {Winter}.
\newblock A resource framework for quantum shannon theory.
\newblock \emph{IEEE Trans. Inf. Theory}, 54\penalty0 (10):\penalty0
  4587--4618, Oct 2008.

\bibitem[Dowling and Milburn(2003)]{DowlingMilburn:03p}
J.~P. Dowling and G.~J. Milburn.
\newblock Quantum technology: the second quantum revolution.
\newblock \emph{Philos. Trans. Royal Soc. London. Series A: Math., Phys. and
  Eng. Sciences}, 361\penalty0 (1809):\penalty0 1655--1674, 2003.

\bibitem[Dupuis(2008)]{Dupuis:08a}
F.~Dupuis.
\newblock Coding for quantum channels with side information at the transmitter.
\newblock \emph{arXiv preprint arXiv:0805.3352}, 2008.

\bibitem[{Dupuis}(2009)]{Dupuis:09c}
F.~{Dupuis}.
\newblock The capacity of quantum channels with side information at the
  transmitter.
\newblock In \emph{Proc. IEEE Int. Symp. Inf. Theory (ISIT'2009)}, pages
  948--952, June 2009.

\bibitem[El~Gamal and Kim(2011)]{ElGamalKim:11b}
A.~El~Gamal and Y.~Kim.
\newblock \emph{Network Information Theory}.
\newblock Cambridge University Press, 2011.

\bibitem[Gel'fand and Pinsker(1980)]{GelfandPinsker:80p}
S.~I. Gel'fand and M.~S. Pinsker.
\newblock Coding for channel with random parameters.
\newblock \emph{Probl. Control Inform. Theory}, 9\penalty0 (1):\penalty0
  19--31, Jan 1980.

\bibitem[Goldsmith et~al.(2009)Goldsmith, Jafar, Maric, and
  Srinivasa]{GJMS:09p}
A.~Goldsmith, S.~A. Jafar, I.~Maric, and S.~Srinivasa.
\newblock Breaking spectrum gridlock with cognitive radios: An information
  theoretic perspective.
\newblock \emph{Proc. of the IEEE}, 97\penalty0 (5):\penalty0 894--914, May
  2009.

\bibitem[{Gyongyosi} et~al.(2018){Gyongyosi}, {Imre}, and
  {Nguyen}]{GyongyosiImreNguyen:18p}
L.~{Gyongyosi}, S.~{Imre}, and H.~V. {Nguyen}.
\newblock A survey on quantum channel capacities.
\newblock \emph{IEEE Commun. Surveys Tutorials}, 20\penalty0 (2):\penalty0
  1149--1205, 2018.

\bibitem[Hastings(2009)]{Hastings:09p}
M.~B. Hastings.
\newblock Superadditivity of communication capacity using entangled inputs.
\newblock \emph{Nature Physics}, 5\penalty0 (4):\penalty0 255, March 2009.

\bibitem[Haykin(2005)]{Haykin:05p}
S.~Haykin.
\newblock Cognitive radio: brain-empowered wireless communications.
\newblock \emph{IEEE J. selected areas in communications}, 23\penalty0
  (2):\penalty0 201--220, Feb 2005.

\bibitem[Heegard and Gamal(1983)]{HeegardElGamal:83p}
C.~Heegard and A.~E. Gamal.
\newblock On the capacity of computer memory with defects.
\newblock \emph{IEEE Trans. Inf. Theory}, 29\penalty0 (5):\penalty0 731--739,
  Sep 1983.

\bibitem[{Holevo}(1998)]{Holevo:98p}
A.~S. {Holevo}.
\newblock The capacity of the quantum channel with general signal states.
\newblock \emph{IEEE Trans. Inf. Theory}, 44\penalty0 (1):\penalty0 269--273,
  Jan 1998.

\bibitem[Holevo(2002)]{Holevo:02p}
A.~S. Holevo.
\newblock On entanglement-assisted classical capacity.
\newblock \emph{J. Math. Phys.}, 43\penalty0 (9):\penalty0 4326--4333, 2002.

\bibitem[{Hsieh} and {Watanabe}(2016)]{HsiehWatanabe:16p}
M.~{Hsieh} and S.~{Watanabe}.
\newblock Channel simulation and coded source compression.
\newblock \emph{IEEE Trans. Inf. Theory}, 62\penalty0 (11):\penalty0
  6609--6619, Nov 2016.

\bibitem[{Hsieh} et~al.(2008){Hsieh}, {Devetak}, and
  {Winter}]{HsiehDevetakWinter:08p}
M.~{Hsieh}, I.~{Devetak}, and A.~{Winter}.
\newblock Entanglement-assisted capacity of quantum multiple-access channels.
\newblock \emph{IEEE Trans. Inf. Theory}, 54\penalty0 (7):\penalty0 3078--3090,
  July 2008.

\bibitem[Imre and Gyongyosi(2012)]{ImreGyongyosi:12b}
S.~Imre and L.~Gyongyosi.
\newblock \emph{Advanced quantum communications: an engineering approach}.
\newblock John Wiley \& Sons, 2012.

\bibitem[Jafar(2006)]{Jafar:06p}
S.~Jafar.
\newblock Capacity with causal and noncausal side information: A unified view.
\newblock \emph{IEEE Trans. Inf. Theory}, 52\penalty0 (12):\penalty0
  5468--5474, Dec 2006.

\bibitem[Jouguet et~al.(2013)Jouguet, Kunz-Jacques, Leverrier, Grangier, and
  Diamanti]{JKLGD:13p}
P.~Jouguet, S.~Kunz-Jacques, A.~Leverrier, P.~Grangier, and E.~Diamanti.
\newblock Experimental demonstration of long-distance continuous-variable
  quantum key distribution.
\newblock \emph{Nature Photonics}, 7\penalty0 (5):\penalty0 378, 2013.

\bibitem[Keshet et~al.(2007)Keshet, Steinberg, and
  Merhav]{KeshetSteinbergMerhav:07n}
G.~Keshet, Y.~Steinberg, and N.~Merhav.
\newblock Channel coding in the presence of side information.
\newblock \emph{Foundations and Trends in Communications and Information
  Theory}, 4\penalty0 (6):\penalty0 445--586, Jan 2007.

\bibitem[Khanian and Winter(2018)]{KhanianWinter:18a}
Z.~B. Khanian and A.~Winter.
\newblock Distributed compression of correlated classical-quantum sources or:
  the price of ignorance.
\newblock \emph{\textup{\texttt{arXiv:1811.09177}}}, 2018.

\bibitem[Khanian and Winter(2019{\natexlab{a}})]{KhanianWinter:19a}
Z.~B. Khanian and A.~Winter.
\newblock Entanglement-assisted quantum data compression.
\newblock \emph{\textup{\texttt{arXiv:1901.06346}}}, 2019{\natexlab{a}}.

\bibitem[Khanian and Winter(2019{\natexlab{b}})]{KhanianWinter:19c}
Z.~B. Khanian and A.~Winter.
\newblock Entanglement-assisted quantum data compression.
\newblock In \emph{Proc. IEEE Int. Symp. Inf. Theory (ISIT'2019)}, pages
  1147--1151, Paris, France, July 2019{\natexlab{b}}.

\bibitem[Khanian and Winter(2019{\natexlab{c}})]{KhanianWinter:19c2}
Z.~B. Khanian and A.~Winter.
\newblock Distributed compression of correlated classical-quantum sources or:
  the price of ignorance.
\newblock In \emph{Proc. IEEE Int. Symp. Inf. Theory (ISIT'2019)}, pages
  1152--1156, Paris, France, July 2019{\natexlab{c}}.

\bibitem[{Khanmohammadi} et~al.(2015){Khanmohammadi}, {Enne}, {Hofbauer}, and
  {Zimmermanna}]{KEHZ:15p}
A.~{Khanmohammadi}, R.~{Enne}, M.~{Hofbauer}, and H.~{Zimmermanna}.
\newblock A monolithic silicon quantum random number generator based on
  measurement of photon detection time.
\newblock \emph{IEEE Photon. J.}, 7\penalty0 (5):\penalty0 1--13, Oct 2015.

\bibitem[Kitaev(1997)]{Kitaev:97b}
A.~Y. Kitaev.
\newblock Quantum error correction with imperfect gates.
\newblock In \emph{Quantum Communication, Computing, and Measurement}, pages
  181--188. Springer, 1997.

\bibitem[Kuznetsov and Tsybakov(1974)]{KuznetsovTsybakov:74p}
A.~V. Kuznetsov and B.~S. Tsybakov.
\newblock Coding in a memory with defective cells.
\newblock \emph{Problemy peredachi informatsii}, 10\penalty0 (2):\penalty0
  52--60, 1974.

\bibitem[Liao et~al.(2017)Liao, Guo, and Huang]{LiaoGuoHuang:17p}
Q.~Liao, Y.~Guo, and D.~Huang.
\newblock Cancelable remote quantum fingerprint templates protection scheme.
\newblock \emph{Chinese Phys. B}, 26\penalty0 (9):\penalty0 090302, 2017.

\bibitem[Lloyd(1997)]{Loyd:97p}
S.~Lloyd.
\newblock Capacity of the noisy quantum channel.
\newblock \emph{Phys. Rev. A}, 55\penalty0 (3):\penalty0 1613, March 1997.

\bibitem[{Luo} and {Devetak}(2009)]{LuoDevetak:09p}
Z.~{Luo} and I.~{Devetak}.
\newblock Channel simulation with quantum side information.
\newblock \emph{IEEE Trans. Inf. Theory}, 55\penalty0 (3):\penalty0 1331--1342,
  March 2009.

\bibitem[Moulin and O'Sullivan(2003)]{MoulinOsullivan:03p}
P.~Moulin and J.~A. O'Sullivan.
\newblock Information-theoretic analysis of information hiding.
\newblock \emph{IEEE Trans. Inf. Theory}, 49\penalty0 (3):\penalty0 563--593,
  Mar 2003.

\bibitem[Nielsen and Chuang(2002)]{NielsenChuang:02b}
M.~A. Nielsen and I.~Chuang.
\newblock Quantum computation and quantum information, 2002.

\bibitem[Qian and Zhang(2018)]{QianZhan:18p}
J.~Qian and L.~Zhang.
\newblock On mds linear complementary dual codes and entanglement-assisted
  quantum codes.
\newblock \emph{Designs, Codes and Cryptography}, 86\penalty0 (7):\penalty0
  1565--1572, 2018.

\bibitem[Schumacher(1995)]{Schumacher:95p}
B.~Schumacher.
\newblock Quantum coding.
\newblock \emph{Phys. Rev. A}, 51\penalty0 (4):\penalty0 2738, 1995.

\bibitem[Schumacher and Nielsen(1996)]{SchumacherNielsen:96p}
B.~Schumacher and M.~A. Nielsen.
\newblock Quantum data processing and error correction.
\newblock \emph{Phys. Rev. A}, 54\penalty0 (4):\penalty0 2629, 1996.

\bibitem[Schumacher and Westmoreland(1997)]{SchumacherWestmoreland:97p}
B.~Schumacher and M.~D. Westmoreland.
\newblock Sending classical information via noisy quantum channels.
\newblock \emph{Phys. Rev. A}, 56\penalty0 (1):\penalty0 131, July 1997.

\bibitem[Shannon()]{Shannon:48p}
C.~Shannon.
\newblock A mathematical theory of communication.
\newblock \emph{Bell Syst. Tech. J}, 27:\penalty0 379--423, 623--656, Jul 1948.

\bibitem[Shannon(1958)]{Shannon:58p}
C.~E. Shannon.
\newblock Channels with side information at the transmitter.
\newblock \emph{IBM J. Res. Dev.}, 2\penalty0 (4):\penalty0 289--293, Oct 1958.

\bibitem[Shirokov(2012)]{Shirokov:12p}
M.~E. Shirokov.
\newblock Conditions for coincidence of the classical capacity and
  entanglement-assisted capacity of a quantum channel.
\newblock \emph{Problems. Inform. Transm.}, 48\penalty0 (2):\penalty0 85--101,
  2012.

\bibitem[Shor(2002{\natexlab{a}})]{Shor:02l}
P.~W. Shor.
\newblock The quantum channel capacity and coherent information.
\newblock In \emph{Lecture notes, MSRI Workshop Quant. Comput.},
  2002{\natexlab{a}}.

\bibitem[Shor(2002{\natexlab{b}})]{Shor:02p}
P.~W. Shor.
\newblock Additivity of the classical capacity of entanglement-breaking quantum
  channels.
\newblock \emph{J. Math. Phys.}, 43\penalty0 (9):\penalty0 4334--4340, May
  2002{\natexlab{b}}.

\bibitem[Smith and Yard(2008)]{SmithYard:08p}
G.~Smith and J.~Yard.
\newblock Quantum communication with zero-capacity channels.
\newblock \emph{Science}, 321\penalty0 (5897):\penalty0 1812--1815, 2008.

\bibitem[Somekh-Baruch et~al.(2008)Somekh-Baruch, Shamai, and
  Verd{\'u}]{BaruchShamaiVerdu:08c}
A.~Somekh-Baruch, S.~Shamai, and S.~Verd{\'u}.
\newblock Cognitive interference channels with state information.
\newblock In \emph{Proc. IEEE Int. Symp. Inf. Theory (ISIT'2008)}, pages
  1353--1357, Toronto, Canada, July 2008.

\bibitem[Steinberg and Merhav(2001)]{SteinbergMerhav:01p}
Y.~Steinberg and N.~Merhav.
\newblock Identification in the presence of side information with application
  to watermarking.
\newblock \emph{IEEE Trans. Inf. Theory}, 47\penalty0 (4):\penalty0 1410--1422,
  May 2001.

\bibitem[{Warsi} and {Coon}(2017)]{WarsiCoon:17p}
N.~A. {Warsi} and J.~P. {Coon}.
\newblock Coding for classical-quantum channels with rate limited side
  information at the encoder: information-spectrum approach.
\newblock \emph{IEEE Trans. Inf. Theory}, 63\penalty0 (5):\penalty0 3322--3331,
  May 2017.

\bibitem[Wilde(2017)]{Wilde:17b}
M.~M. Wilde.
\newblock \emph{Quantum information theory}.
\newblock Cambridge University Press, 2 edition, 2017.

\bibitem[{Wilde} et~al.(2014){Wilde}, {Hsieh}, and
  {Babar}]{WildeHsiehBabar:14p}
M.~M. {Wilde}, M.~{Hsieh}, and Z.~{Babar}.
\newblock Entanglement-assisted quantum turbo codes.
\newblock \emph{IEEE Trans. Inf. Theory}, 60\penalty0 (2):\penalty0 1203--1222,
  Feb 2014.

\bibitem[Winter(2016)]{Winter:16p}
A.~Winter.
\newblock Tight uniform continuity bounds for quantum entropies: conditional
  entropy, relative entropy distance and energy constraints.
\newblock \emph{Commun. in Math. Phys.}, 347\penalty0 (1):\penalty0 291--313,
  2016.

\bibitem[Wyner and Ziv(1976)]{WynerZiv:76p}
A.~Wyner and J.~Ziv.
\newblock The rate-distortion function for source coding with side information
  at the decoder.
\newblock \emph{IEEE Trans. Inf. Theory}, 22\penalty0 (1):\penalty0 1--10, Jan
  1976.

\bibitem[{Yard} and {Devetak}(2009)]{YardDevetak:09p}
J.~T. {Yard} and I.~{Devetak}.
\newblock Optimal quantum source coding with quantum side information at the
  encoder and decoder.
\newblock \emph{IEEE Trans. Inf. Theory}, 55\penalty0 (11):\penalty0
  5339--5351, Nov 2009.

\bibitem[Yin et~al.(2017)Yin, Cao, Li, Liao, Zhang, Ren, Cai, Liu, Li, Dai, Li,
  Lu, Gong, Xu, Li, Li, Yin, Jiang, Li, Jia, Ren, He, Zhou, Zhang, Wang, Chang,
  Zhu, Liu, Chen, Lu, Shu, Peng, Wang, and Pan]{YCLZRC:17p}
J.~Yin, Y.~Cao, Y.~H. Li, S.~K. Liao, L.~Zhang, J.~G. Ren, W.~Q. Cai, W.~Y.
  Liu, B.~Li, H.~Dai, G.~B. Li, Q.~M. Lu, Y.~H. Gong, Y.~Xu, S.~L. Li, F.~Z.
  Li, Y.~Y. Yin, Z.~Q. Jiang, M.~Li, J.~J. Jia, G.~Ren, D.~He, Y.~L. Zhou,
  X.~X. Zhang, N.~Wang, X.~Chang, Z.~C. Zhu, N.~L. Liu, Y.~A. Chen, C.~Y. Lu,
  R.~Shu, C.~Z. Peng, J.~Y. Wang, and J.~W. Pan.
\newblock Satellite-based entanglement distribution over 1200 kilometers.
\newblock \emph{Science}, 356\penalty0 (6343):\penalty0 1140--1144, 2017.

\bibitem[Zhang et~al.(2017)Zhang, Ding, Sheng, Zhou, Shi, and Guo]{ZDSZSG:17p}
W.~Zhang, D.~S. Ding, Y.~B. Sheng, L.~Zhou, B.~S. Shi, and G.~C. Guo.
\newblock Quantum secure direct communication with quantum memory.
\newblock \emph{Phys. Rev. Lett.}, 118\penalty0 (22):\penalty0 220501, 2017.

\end{thebibliography}

\end{document}